\input harvmac
\catcode`@=12
\baselineskip12pt
\noblackbox
\def\undertext#1{$\underline{\smash{\hbox{#1}}}$}


\global\newcount\propno \global\propno=1
\def\prop#1{\xdef #1{\secsym\the\propno}\writedef{#1\leftbracket#1}%
{\bf\the\secno.}{\bf\the\propno.}
\global\advance\propno by1
\proplabeL#1}

\def\proplabeL#1{\leavevmode\vadjust{\llap{\smash 
   {\line{{\escapechar=` \hfill\llap{\sevenrm\hskip.03in\string#1$\;\;$}}}}}}}

\def\undertext#1{$\underline{\smash{\hbox{#1}}}$}

\def\IP{{\bf P}}
\def\IR{{\bf R}}
\def\IC{{\bf C}}
\def\IQ{{\bf Q}}
\def\ZZ{{\bf Z}}
\def\pd{\partial}
\def\D{{\Delta}}

\def\Ds{{\Delta^*}}

\def\({ \left(  }
\def\){ \right) }
\def\S{\Sigma}
\def\SD{{\Sigma(\D)}}
\def\SDs{{\Sigma(\Ds)}}

\def\bns#1{\bar\nu_{#1}^*}
\def\ns#1{\nu_{#1}^*}

\def\rms{{\rm -}}
\def\Vol{{\rm Vol}}
\def\ch{{\rm ch}}
\def\Todd{{\rm Todd}}
\def\frac#1#2{{#1 \over #2}}

\def\mapright#1{\smash{\mathop{\longrightarrow}\limits^{#1}}} 
\def\mapdown#1{\Big\downarrow\rlap{$\vcenter{\hbox{$\scriptstyle#1$}}$}}

%
%
\def\AGM{[1]}
\def\refAGM{P.S. Aspinwall, B.R. Greene and D.R. Morrison,
  {\it Multiple mirror manifolds and topology change in string theory},
  Phys. Lett. B303 (1993) 249-259. }
\def\Bat{[2]}
\def\refBat{
  V.V. Batyrev, {\it Dual polyhedra and mirror symmetry for Calabi-Yau 
  hypersurfaces in toric varieties}, J. Alg. Geom. 3(1994), 493-535.}
\def\BO{[3]}
\def\refBO{A. Bondal and D. Orlov, {\it Semiorthogonal decompositions for 
  algebraic varieties}, alg-geom/9506012.}
\def\Bri{[4]}
 \def\refBri{ T. Bridgeland, {\it Equivalences of triangulated categories 
  and Fourier-Mulai transforms}, Bull. London Math. Society {\bf 31}(1999), 
  25-34, math.AG/9809114. } 
\def\BDoug{[5]}
\def\refBDoug{ I. Brunner, M.R. Douglas, A. Lawrence and C. R\"omelsberger, 
            {D-branes on the Quintic  }, hep-th/9906200.}
\def\CdGP{[6]}
\def\refCdGP{ 
  P. Candelas, X.C. de la Ossa, P.S. Green, and L.Parkes, 
  {\it A pair of Calabi-Yau manifolds as an exactly soluble 
  superconformal theory}, Nucl.Phys. B356(1991), 21-74. }
\def\CdTwoI{[7]}
\def\refCdTwoI{
  P. Candelas, X.C. de la Ossa, A. Font, S. Katz and D.R. Morrison, 
  {\it Mirror symmetry for two parameter models - I}, Nucl. Phys. B {\bf 416}
  (1994), 481-538, hep-th/9308083.}
\def\CdTwoII{[8]}
\def\refCdTwoII{
  P. Candelas, X.C. de la Ossa, A. Font, S. Katz and D.R. Morrison, 
  {\it Mirror symmetry for two parameter models - II}, Nucl. Phys. B {\bf 429}
  (1994), 626-674, hep-th/9403187.}
\def\CKYZ{[9]}
\def\refCKYZ{ 
  T.-T. Chiang, A. Klemm, S.-T. Yau and E. Zaslow, 
  {\it Local Mirror Symmetry: Calculations and Interpretations}, 
  hep-th/9903053.}
\def\Cox{[10]}
\def\refCox{  D. Cox, 
   {\it The homogeneous coordinate ring of a toric variety}, J. Alg. Geom. 
   {\bf 4}(1995)17-50, alg-geom/9210008.}
\def\DR{[11]}
\def\refDR{D. Diaconescu and C. R\"omelsberger, {\it D-branes 
   and bundles on elliptic fibrations}, hep-th/9910172. }
\def\Dol{[12]}
\def\refDol{ I.V. Dolgachev,{\it Mirror symmetry for lattice polarized K3
 surfaces}, Algebraic geometry, 4. J. Math. Sci. 81 (1996), no. 3, 2599--2630.}
\def\DFR{[13]}
\def\refDFR{ M. Douglas, B. Fiol and C. R\"omelsberger, {\it 
    The spectrum of BPS branes on a noncompact Calabi-Yau}, hep-th/0003263.}
\def\Ful{[14]}
\def\refFul{ W. Fulton, {\it Introduction to Toric Varieties},
  Ann. of Math. Studies 131, Princeton University Press,
  Princeton, New Jersey, 1993. }
\def\Gi{[15]}
\def\refGi{ A.B. Givental, {\it Equivariant Gromov-Witten invariants}, 
  Internat. Math. Res. Notices 13 (1996) 613-663.} 
\def\HIV{[16]}
\def\refHIV{ K. Hori, A. Iqbal and C. Vafa, 
  {\it D-branes and mirror symmetry}, hep-th/0005247.}
\def\Horja{[17]}
 \def\refHorja{ R.P. Horja, 
        {\it Hypergeometric Functions and Mirror Symmetry in Toric Varieties}, 
         math.AG/9912109. }
\def\Hos{[18]}
\def\refHos{ S. Hosono, {\it GKZ systems, Gr\"obner fans and moduli spaces 
  of Calabi-Yau hypersurfaces} 
  in ``Topological Field Theory, Primitive Forms and Related Topics''
  Progress in Mathematics, Birkh\"auser, pp.239-265, alg-geom/9707003. } 
\def\HKTYI{[19]}
\def\refHKTYI{ 
  S. Hosono, A. Klemm, S. Theisen, and S.-T. Yau, 
  {\it Mirror symmetry, mirror map and applications to Calabi-Yau 
  hypersurfaces},  
  Commun. Math. Phys. 167(1995), 301-350, hep-th/9308122. } 
\def\HKTYII{[20]}
\def\refHKTYII{ S. Hosono, A. Klemm, S. Theisen, and S.-T. Yau, 
  {\it Mirror symmetry, mirror map and applications to complete 
  intersection Calabi-Yau spaces},  
  Nucl. Phys. B433 (1995) 501-554, hep-th/9406055.  }
\def\HLY{[21]}
\def\refHLY{ S. Hosono, B.H. Lian, and S.-T. Yau, {\it  
  GKZ-Generalized hypergeometric systems in mirror symmetry 
  of Calabi-Yau hypersurfaces},
  Commun. Math. Phys. 182 (1996) 535-577, alg-geom/9511001.  }
\def\HLYII{[22]}
\def\refHLYII{ S. Hosono, B.H. Lian, and S.-T. Yau, {\it 
  Maximal degeneracy points of GKZ systems}, 
  J. Amer. Math. Soc. 10(1997), 427-443, alg-geom/9603014. }
\def\HLYm{[23]}
 \def\refHLYm{ S. Hosono, B. H. Lian and S.-T. Yau, 
        {\it Type IIA monodromy of Calabi-Yau manifolds}, 
        1997, (unpublished). }
\def\KLLW{[24]}
\def\refKLLW{ P. Kaste, W. Lerche, C.A. L\"utken and J. Walcher, 
   {\it D-branes on K3-fibrations}, hep-th/9912147.}
\def\KLM{[25]}
\def\refKLM{ A. Klemm, W. Lerche and P. Mayr, {\it K3 fibrations 
  and heterotic type II string duality}, Phys. Lett. B357 (1995)313-322, 
  hep-th/9506112.}
\def\KMV{[26]}
\def\refKMV{ 
   A. Klemm, P. Mayr and C. Vafa,
   {\it BPS States of Exceptional Non-Critical Strings}, 
   in the proceedings of the conference "Advanced Quantum Field Theory'' 
   (in memory of Claude Itzykson) hep-th/9607139. }
\def\KT{[27]}
 \def\refKT{ A.Klemm and S.Theisen, 
          {\it Considerations of one modulus Calabi-Yau Compactifications: 
           Picard-Fuchs equations, K\"ahler potentials and mirror maps}, 
           Nucl. Phys. B389(1993)153-180. }
\def\Kob{[28]}
\def\refKob{ M. Kobayashi, {\it Duality of weights, mirror
   symmetry and Arnold's strange duality}, alg-geom/9502004.} 
\def\KoI{[29]}
\def\refKoI{ M. Kontsevich, {\it Homological algebra of mirror symmetry}, 
  in Proceedings of the International Congress of Mathematicians, 
  Vol. 1, 2 (Z\"uich, 1994) 120-139, Birkh\"auser, Basel, 1995.} 
\def\KoII{[30]}
\def\refKoII{ M. Kontsevich, lectures, 1996, (unpublished) } 
\def\LMW{[31]}
\def\refLMW{ W. Lerche, P. Mayr and N.P. Warner, 
  {\it Non-critical strings, del Pezzo singularities 
   and Seiberg-Witten curves}, Nucl.Phys. B499 (1997) 125-148, hep-th/9612085.}
\def\LLY{[32]}
\def\refLLY{ B.H. Lian, K. Liu and S.-T. Yau, 
  {\it Mirror principle I, II},  Asian J. Math. 1 (1997), no.4,729-763, and   
  Asian J. Math. 3 (1999), no. 1, 109--146, alg-geom/9712011,  math/9905006. }
\def\LY{[33]}
\def\refLY{  B.H. Lian and S.-T. Yau, 
  {\it Arithmetic properties of the mirror map and quantum coupling}, 
  Commun. Math. Phys. 176 (1996) 163-192, hep-th/9411234.}
\def\Mor{[34]}
\def\refMor{ D. R. Morrison, {\it Picard-Fuchs equations and mirror maps 
  for hypersurfaces} in {\it Essays on Mirror Manifolds} ed. S.-T.Yau, International Press, Hong Kong 1992, alg-geom/9202026.} 
\def\MVI{[35]}
\def\refMVI{ D.R. Morrison and C. Vafa, {\it Compactifications of 
   F-Theory on Calabi--Yau Threefolds -- I}, Nucl. Phys. B473 (1996) 74-92.} 
\def\MVII{[36]}
\def\refMVII{ D.R. Morrison and C. Vafa, {\it Compactifications of 
   F-Theory on Calabi--Yau Threefolds -- II}, Nucl. Phys. B476 (1996) 437-469.}
\def\MukaiI{[37]}
\def\refMukaiI{ S. Mukai, {\it Duality between $D(X)$ and $D(\tilde X)$ 
  with its application to Picard sheaves}, Nagoya Math. J. {\bf 81}, 153-175.}
\def\MukaiII{[38]}
\def\refMukaiII{ S. Mukai, {\it On the moduli spaces of bundles on a 
   K3 surface I}, Vector bundles on algebraic varieties (Bombay, 1984), 
   Tata. Inst. Fund. Res., Bombay, 1987, 341-413.}
\def\Oda{[39]}
\def\refOda{T. Oda,{\it Convex bodies and Algebraic Geometry, An
  Introduction to the Theory of Toric Varieties},
  A Series of Modern Surveys in Mathematics,
  Springer-Verlarg New York, 1985.} 
\def\OOY{[40]}
\def\refOOY{ H. Ooguri, Y. Oz and Z. Yin, {\it D-branes on Calabi-Yau spaces 
  and their mirrors}, Nucl.Phys.B477 (1996), 407-430, hep-th/9606112.}
\def\Sch{[41]}
\def\refSch{ E. Scheidegger, {\it D-branes on one- and two-parameter 
Calabi-Yau hypersurfaces}, JHEP 0004 (2000) 003, hep-th/9912188.}
\def\Sti{[42]}
\def\refSti{J. Stienstra, {\it Resonant hypergeometric systems and mirror 
symmetry}, in {\it Integrable systems and algebraic geometry 
(Kobe/Kyoto 1997)}, World Sci. Publishing, River Edge, NJ, 1998, 412-452, 
alg-geom/9711002.}
\def\Str{[43]}
\def\refStr{ A. Strominger, {\it Special Geometry}, 
  Commun. Math. Phys. 133 (1990) 163-180. }
\def\SYZ{[44]}
\def\refSYZ{ A. Strominger, S.-T. Yau and E. Zaslow, 
     {\it Mirror symmetry is T-Duality}, Nucl. Phys. B479 (1996) 243-259.}
\def\Vafa{[45]}
\def\refVafa{ C. Vafa, {\it Extending Mirror Conjecture to Calabi-Yau 
          with Bundles}, hep-th/9804131.}

\rightline{June, 2000}
\vskip0.5cm
\centerline{\bf 
Local Mirror Symmetry and Type IIA Monodromy of 
Calabi-Yau manifolds }

\vskip1cm

\centerline{
Shinobu Hosono$^\dagger$  
\footnote{}{\hskip-0.8cm email:
$^\dagger$  hosono@ms.u-tokyo.ac.jp}  }

\bigskip
\centerline{ Graduate School of Mathematical Sciences  }
\centerline{      University of Tokyo   }
\centerline{   Komaba 3-8-1, Meguro-ku  } 
\centerline{   Tokyo 153-8914,  Japan   }

\vskip1cm
\noindent {\bf Abstract:} 
We propose a monodromy invariant pairing $K_{hol}(X) \otimes H_3(X^\vee,\ZZ) 
\rightarrow \IQ$ for a mirror pair of Calabi-Yau manifolds, $(X,X^\vee)$. 
This pairing is utilized implicitly in the previous calculations of the 
prepotentials for Gromov-Witten invariants. After identifying the pairing 
explicitly we interpret some hypergeometric series from the viewpoint of 
homological mirror symmetry due to Kontsevich. Also we consider the local 
mirror symmetry limit to del Pezzo surfaces in Calabi-Yau 3-folds.

\newsec{ Introduction } 

In 1994, Kontsevich proposed a homological mirror symmetry{\KoI} for 
a mirror pair of Calabi-Yau 3 folds. It is conjectured that there exists 
a categorical equivalence between the (bounded) derived category $D(X)$ 
of $X$ and Fukaya's $A_\infty$ category of Lagrangian submanifolds for 
the mirror $X^\vee$. The object of Fukaya's $A_\infty$ category is a (special) 
Lagrangian submanifold with flat $U(1)$ bundles on it and undergoes 
the monodromy transformations when we consider the family 
of manifolds $X^\vee$. On the other hand the object of the derived 
category $D(X)$ is a complex of coherent sheaves on $X$. Under the 
(conjectural) categorical equivalence between $D(X)$ and Fukaya's 
$A_\infty$ category the monodromy of the 3-cycles is mapped to 
certain automorphisms in $D(X)$, Fourier-Mukai transformations. 
To write this automorphism as well as to study the equivalence 
of the category explicitly are important problems 
toward the complete understanding of the mirror symmetry. 

About the possible form of the automorphisms in $D(X)$ there 
have been a proposal by Kontsevich{\KoII}, and his idea has 
been pushed forward recently by Horja{\Horja}. They propose  
certain Fourier-Mukai transformations on $D(X)$ to reproduce the 
corresponding monodromy actions on the period integrals 
and compare their effects on (the D-brane charges) 
$H^{even}(X,\IC)$, i.e., the cohomology 
ring of even degrees over the coefficient $\IC$. 
In this paper we remark a natural integral structure for the 
D-brane charges which was implicit in the previous 
calculations{\HKTYII}{\HLY}{\HLYm}.  Also we will identify 
the generalization by Horja coincides with the local mirror 
symmetry in physics{\LMW}{\CKYZ} and consider the cases $\IP^2$ and 
del Pezzo surfaces $Bl_6, Bl_7, Bl_8$ in detail.  

The integral structure above comes from 
that of the K-group $K(X)$ by the Chern character homomorphism. As for 
the K-group, we consider the topological $K(X)$, i.e., the set of 
vector bundles on $X$ divided by a certain relation 
(the stable equivalence). Although not all 
elements in $D(X)$ (, or coherent sheaves on $X$,) correspond to 
vector bundles on $X$, we assume all elements in $D(X)$ 
may be written as a complex of vector bundles and have their images 
in $K(X)$. Throughout this paper we write by $K(X)_{hol}$ the 
image of $D(X)$ in $K(X)$ under this assumption.

Main result of this paper is a proposal (``Theorem'' 2); {\it 
There exists a monodromy invariant pairing 
\eqn\IinvP{
I: K_{hol}(X) \otimes H_3(X^\vee,\ZZ) \rightarrow  \IQ
}
for the deformation family of Calabi-Yau manifolds $X^\vee$ and its 
mirror manifold $X$. }  We derive this results from Proposition 1 
which has been verified explicitly for abundance of examples to 
establish the mirror symmetry conjecture{\CdGP}. For some special cases 
the mirror symmetry conjecture has been proved in {\Gi}{\LLY}. 

The homological mirror symmetry was also considered by Vafa{\Vafa} to extend  
the mirror symmetry conjecture{\CdGP}. In ref.{\Vafa} the yukawa coupling 
on the complex structure moduli space was modified to incorporate the moduli 
space of vector bundles on Calabi-Yau manifolds, and it was conjectured that 
the generalized formula enumerates open strings with boundary 
on a special Lagrangian 3-cycle in the mirror Calabi-Yau manifold.

The organization of this paper is as follows: In section 2, we will 
summarize the mirror symmetry of hypersurfaces in toric varieties 
following{\HLY}{\Hos}. In section 3 we will define type IIA monodromy 
and a natural integral symplectic structure implicit in the calculations 
summarized in section 2.  In section 4, we will consider the monodromy 
associated to the local mirror symmetry to $\IP^2$ and del Pezzo surfaces 
$Bl_6,Bl_7, Bl_8$ and interpret their integral structure in language 
of coherent sheaves.  In the final section, we will address to the 
recent developments in string theory which are closely related to the 
content of this paper.

\vskip0.5cm
\noindent
{\bf Acknowledgement}: The author would like to thank B.H. Lian and S.-T. Yau 
for their collaboration for the unpublished work {\HLYm}, the preliminary 
results there are the starting point of this work. 
He also would like to thank H. Grosse, M. Kreuzer and S. Theisen  for 
their kind invitation to 
``Duality, String Theory and M-Theory'' at Ervin Sr\"odinger Institute 
(March,2000), where part of this work has been done. He also thanks 
A. Tsuchiya, A. Klemm, K. Yoshioka, E. Zaslow for discussions and C. Vafa 
for drawing his attention to the generalized mirror symmetry 
conjecture{\Vafa}. 
This work is supported in part by Grant-in Aid for Science Research 
A11740006.

\vfill\eject

\newsec{ GKZ hypergeometric series and Prepotential --- Review}

In this section we summarize general results about the prepotential 
of Calabi-Yau hypersurfaces in toric varieties obtained in {\HKTYII}{\HLY}. 
For simplicity we will mainly focus on hypersurfaces, however generalizations 
to complete intersections are straightforward(, see {\HKTYII} for example). 

\subsec{ GKZ hypergeometric series and large complex structure limit}

Let us consider a mirror pair of Calabi-Yau hypersurfaces $(X_\D,X_\Ds)$ 
described by (four dimensional) reflexive polytopes $(\D,\Ds)$ {\Bat}. 
We denote the lattice points of the polytope $\Ds$ by $\ns{k} \; 
(k=0,\cdots, p)$ with a convention $\ns{0}=\vec 0$. Then the defining 
equation of $X_\Ds$ is given by 
$$
f(a)=\sum_{i=0}^p a_i X^{\ns{i}} \;\;,
$$
in terms of the torus coordinate $X_1, \cdots, X_4$ of 
$(\IC^*)^4 \subset \IP_{\SD}$. The complex parameters $(a_0, \cdots, a_p) 
\in (\IC^*)^{p+1}$ represents the deformations of the defining equation.  
In nice situations, to which our attention will mainly be restricted, the 
monomial deformations of the defining equation provide the complex structure 
deformations of $X_{\Ds}$.  We compactify the affine space described by 
$a_i$'s via the secondary fan so that the natural $\IC^*$ actions coming 
from the toric ambient space $\IP_{\SD}$ are respected. The basic object 
we consider in the compactification is the following lattice, so-called 
the lattice of relations among the vertices,
$$
L=\{ \; (l_0,\cdots,l_p) \in \ZZ^{p+1} \; | \; 
  \sum_{i=0}^p l_i \bar\ns{i} = \vec 0 \;,\; \bar \ns{i}=(1,\ns{i}) \;\}.
$$ 
The secondary fan is a rational, polyhedral, complete fan in $L_{\IR}^\vee$ 
for the dual lattice $L^\vee$ and its scalar extension. According to 
the general construction of toric varieties, the complete fan 
defines a toric compactification of the torus ${\rm Hom}_{\ZZ}(L,\IC^*)$, 
which is our compactification of the affine space of $a_i$'s. Among the 
cones in the secondary fan, there is a cone, called Mori cone, whose 
geometric meaning is the dual of the K\"ahler cone of $\IP_{\SDs}$. Let 
us denote the generators of the Mori cone $l^{(1)}, \cdots, l^{(r)} \; 
(r={rk}(H^2(\IP_{\SDs},\ZZ)))$ assuming that it is simplicial for 
simplicity (see {\HLY} for general non-simplicial cases). 
Then the Mori cone $L_{\geq 0} 
= \IR_{\geq0} l^{(1)}+ \cdots + \IR_{\geq0} l^{(r)}$ describes the 
affine chart $U_0={\rm Hom}_{s.g.}(L_{\geq0},\IC)$ with the coordinate 
$x_k=(-1)^{l^{(k)}_0} a^{l^{(k)}} \; (k=1,\cdots,r)$. It has been proved 
in general {\HLYII} that the origin $x_1=\cdots=x_r=0$ provides  a  
large complex structure limit (LCSL){\Mor} and there we have only one 
regular period integral 
\eqn\fundamentalwo{
w_0(x)=\sum_{n \in \ZZ^r_{\geq0}} { \Gamma(1-\sum_k n_k l_0^{(k)}) 
\over \prod_{i=1}^p \Gamma(1+\sum_k n_k l^{(k)}_i) } x^k \;\;,
}
which corresponds to the special Lagrangian three tori cycle{\SYZ}. All 
other period integrals at LCSL contains logarithmic singularities. In 
ref.{\HKTYI} such solutions are generated by classical Frobenius method. 
Before going into the details of the construction of the local 
solutions, we prepare some notations for the cohomology ring 
of the ambient toric variety $\IP_\SDs$. At a first looking, the procedure 
in what follows seems to be simply technical {\it but} it will turn out that 
there are important implications in view of homological mirror symmetry. 

\vskip0.5cm

\subsec{Chow ring of $\IP_\SDs$}

Here we describe the Chow ring of 
the toric variety $\IP_\SDs$ to determine the necessary 
topological data of $X_\D$. 
The Chow ring is an abelian group generated by
divisors, endowed with a commutative, graded ring structure via the
intersection product. In case of toric varieties, it has a simple
description in terms of the invariant divisors $D_i\; (i=0,\cdots,p)$
{\Oda}{\Ful}:
\eqn\chow{
A^*(\IP_\SDs)=\IQ[D_0,D_1,\cdots,D_p]/(SR+I) \;,}
where $SR$ is the Stanley-Reisner ideal for the polytope $\Ds$ and 
$I$ is the ideal generated by linear relations, 
representing rational equivalences among divisors;
\eqn\Ilin{
\sum_{i=0}^p \bns{i} D_i = \vec 0   \;\; . }
We note here a natural non-degenerate pairing $A^1(\IP_\SDs)_\IR \times L_\IR 
\rightarrow \IR$, and thus recognize the dual cone $(L_{\geq0})^\vee$
lies in $A^1(\IP_\SDs)_\IR$. In fact, according to our construction 
of $L_{\geq 0}$, $(L_{\geq0})^\vee$ is the 
K\"ahler cone (,or its refinement,) of the ambient space $\IP_\SDs$. 
We will denote the dual basis to the generators $l^{(1)},\cdots,l^{(r)}$ by  
$J_1,\cdots,J_r$. Then we may write, for example, the classical coupling
$K^{cl}_{abc}$ by 
\eqn\Kcl{
K^{cl}_{abc}:=\int_{\IP_\SDs}[X_\D]\cdot J_a \cdot J_b \cdot J_c \;,}
where $[X_\D]=D_1+\cdots+D_p$ and the symbol $\int_{\IP_\SDs}$ means 
to take the coefficient of the highest degree element of $\chow$ 
with the normalization determined by the requirement that it gives 
the Euler number $\chi(X_\D)$ from the top Chern class $c_n(X_\D)$,( see 
(3.59) of {\HLY} for details). 
The toric part of the even cohomology $H^{even}_{toric}(X_\D,\IQ)$ may
be described by $A^*(\IP_\SDs)_\IR/Ann([X_\D])$, where 
and $Ann([X_\D]):=\{ v \in A^*(\IP_\SDs) \;|\; [X_\D]v=0 \}$.

\subsec{ Prepotential and its asymptotic form}

Now we describe the local solutions about the LCSL.  To this aim 
let us introduce in {\fundamentalwo} the indices $\rho_1,\cdots, \rho_r$;
$$
w_0(x,\rho)=\sum_{n \in \ZZ^r_{\geq 0}} c(n+\rho)x^{x+\rho} \;\;.
$$
The results obtained in {\HKTYII} (or (3.38) of {\HLY}) may be summarized 
into a concise formula using  
the dual bases $J_1, \cdots, J_r$ to $l^{(1)},\cdots, l^{(r)}$ 
for the K\"ahler cone of $\IP_{\SDs}$ and restricting them to the 
hypersurface $X_{\D}$.  Namely the differentials 
with respect to the indices in Frobenius method are summarized by 
$$ 
\eqalign{
&w_0(x,{J \over 2\pi i})=w_0(x) {\bf 1}  \cr
&\;\; + \sum_{a=1}^r w_a(x) J_a 
     + {1\over2!}\sum_{a,b=1}^r w_{ab}(x)J_a J_b 
     + {1\over3!}\sum_{a,b,c=1}^r w_{abc}(x)J_a J_b J_c 
\;\;, \cr} 
$$
where products of $J_k$'s are taken in the cohomology ring 
$H^{even}_{toric}(X_\D,\IQ)$. This way of keeping track of the hypergeometric 
series was first utilized in {\Gi} and later used in {\Sti}. 
We stress here that our definition of the hypergeometric series 
{\fundamentalwo} (or (3.38) of {\HLY}) and that in {\Sti} 
differs in the Gamma factor which normalizes the solutions. 
This difference in the Gamma factor is crucial 
to obtain integral, symplectic basis for the period integrals, although 
both give the solutions of Picard-Fuchs equation.  
Now to describe the formulas more 
concretely, let us introduce a basis ${\bf 1}, J_a, J_b^{(2)}, J^{(3)}$ 
of respective degrees by the property 
$$
({\bf 1}, J^{(3)}) = -1 \;\;,\;\; (J_a,J^{(2)}_b)=\delta_{ab} \;\;,
$$
with $(A,B):=\int_{X_\D} A \wedge B $. From the reasons which will become 
clear soon, we shift the basis slightly to  
introduce a canonical ``symplectic'' basis of $H^{even}_{toric}(X_\D,\IQ)$ by
\eqn\sympIIa{
{\bf 1} \;,\; J_a^{(1)}:=J_a-{c_2(X_\D)\wedge J_a \over 12} \;,\; 
J_b^{(2)} \;,\; J^{(3)} \;\;.}

The above formal expansion of the series may be connected to the 
classical Frobenius method by
\eqn\wsDef{
w_0(x,{J \over 2\pi i}) = w_0(x){\bf 1} + \sum_a D_a^{(1)}w_0(x) J_a^{(1)} 
+ \sum_b D^{(2)}_b w_0(x) J^{(2)}_b + \tilde D^{(3)} w_0(x) J^{(3)} \;,
}
where 
$$
\eqalign{
& D^{(1)}_a = {1\over 2\pi i} \pd_{\rho_a} \;\;,\;\; 
D^{(2)}_b = {1\over 2! (2\pi i)^2}
\sum_{c,d} K^{cl}_{bcd}\pd_{\rho_c}\pd_{\rho_d} \;\;, \cr
\tilde D^{(3)} 
& =-{1\over 3! (2\pi i)^3} \sum_{a,b,c} K^{cl}_{abc} 
\pd_{\rho_a}\pd_{\rho_b}\pd_{\rho_c} - {1\over 2\pi i} \sum_a 
{(c_2(X_\D),J_a) \over 12 } \pd_{\rho_a} \;\;,  \cr
}
$$
and the notation $D^{(1)}_a w_0(x)$, for example, means an operation 
${\rm lim}_{\rho \rightarrow 0} D^{(1)}_a w_0(x,\rho)$.  The form 
of the differential operators $D^{(k)}$ using the classical yukawa 
coupling $K^{cl}_{abc}=\int_{X_\D} J_a\wedge J_b \wedge J_c$ appeared 
in the references {\HKTYII} (see also (3.38) of {\HLY}) and may 
be reproduced directly from 
the formal definitions of $w_0(x,{J \over 2\pi i})$ and the basis 
{\sympIIa}.  In terms of the above hypergeometric series, a closed 
formula for the genus zero prepotential was found to be
\eqn\prepotF{
\eqalign{
F(t)&={1\over2}\({1\over w_0(x)}\)^2 \big\{ w_0(x)\tilde D^{(3)}w_0(x) 
+\sum_a D^{(1)}_a w_0(x) D^{(2)}_a w_0(x) \big\} \;, \cr 
&\qquad t_a = {D^{(1)}_a w_0(x) \over w_0(x) } \;\; (a=1,\cdots,r). \cr}
}
Especially a specific asymptotic form of the prepotential at 
the large radius limit was observed in {\HKTYII} generalizing the results 
for the quintic {\CdGP}, which we reproduce here in the form 
$$
F(t)=\int_{X_\D} {\cal F}(t,J) \;\;,
$$
with ${\cal F}(t,J)$ defined by
\eqn\invF{
{\cal F}(t,J)= 
{1\over6}(t\cdot J)^3 - {c_2(X_\D) \over 24}\wedge (t\cdot J) 
+{\zeta(3) \over 2(2\pi i)^3} c_3(X_\D)  
-{1\over 2} 
\log\(\sum_{n\in \ZZ_{\geq0}^r} 
{c(n+{J \over 2\pi i} ) \over c({J \over 2\pi i})}  x^n \) \;\;,  }
where $t\cdot J =\sum_a t_a J_a$. For general proof of this formula 
we refer to a review article {\Hos} (Proposition 6.7). 
We remark here that the prepotential was introduced to define the 
special K\"ahler geometry on the complex structure moduli space 
using a symplectic basis of the homology group $H_3(X_\Ds,\ZZ)${\Str}. 
Combined this fact with 
our results {\invF}, we may expect that the right 
asymptotic form of the prepotential indicates that the hypergeometric 
series we have arranged about the LCSL in {\wsDef} is very close to 
the integral symplectic basis of period integrals. In the next section, 
we will argue that this is in fact the case relating  the monodromy problem to 
the recent progresses on the Fourier-Mukai transformations on derived category 
by Horja{\Horja}.

\vskip0.5cm

\subsec{ Lattice polarized K3 surfaces }

We may apply the above construction to the lattice polarized K3 
surfaces {\Dol}{\Kob} in parallel ways. The lattice polarized K3 surfaces 
are defined by a lattice $M$ which has the signature $(1,r-1)$ and 
also admit a primitive embedding into the K3 lattice $L_{K3}=E_8(-1)\oplus 
E_8(-1) \oplus H^{\oplus 3}$ where $H$ represents the rank two hyperbolic 
lattice. Given a lattice $M$, the lattice polarized K3 surface, 
$M$-polarized K3 surface, is defined to be the K3 surface whose Picard 
lattice is given by $M$. In a parallel way to Calabi-Yau 3 folds, we 
can construct mirror pair of  K3 surfaces 
for a reflexive pair of three dimensional polytopes $(\D,\Ds)$, thereby we 
obtain the lattice polarized K3 surfaces $X_\D$ and $X_\Ds$ with their 
Picard lattices given by $M_\D$ and $M_\Ds$, respectively. The two lattices 
$M_\D$ and $M_\Ds$ admit primitive embedding into the K3 lattice and allow 
the inclusion 
\eqn\inclusion{
L_{K3} \supset M_\D \oplus H \oplus M_\Ds \;\;,
}
with finite index (, see {\Dol}{\Kob}). 
The orthogonal complement of the Picard lattice 
gives the transcendental lattice of the K3 surface, and gives up to 
the factor $H$ the Picard lattice of the mirror K3 surface. 

Now let us consider the prepotentials for the K3 surface $X_\D$.    
Since it is known that for K3 surfaces there are 
no instanton corrections in the prepotential due to quadratic 
relations (period relations) among period integrals, we may expect 
strict constraints on the integral basis for the period integrals 
of the mirror $X_\Ds$. Namely we should have
$$
F(t)={1\over2} \sum_{a,b} K^{cl}_{ab}  t_a t_b \;\;.
$$
for the prepotential of $X_\D$. With this in mind, 
we arrange the series $w_0(x,{J\over 2\pi i})$ in {\wsDef} by 
\eqn\wsKiii{
w_0(x,{J \over 2\pi i})=w_0(x) {\bf 1}' 
+ \sum_{a=1}^r D^{(1)}_a w_0(x) J^{(1)}_a 
+ \tilde D^{(2)}w_0(x) J^{(2)} \;\;,
}
where 
\eqn\KiiiBasis{
{\bf 1}'={\bf 1} -{c_2(X_\D) \over 24} \;\;,\;\;
J_a^{(1)} = J_a \;\;,\;\; 
J^{(2)}=-{c_2(X_\D) \over 24} 
\;\;, }
and 
\eqn\DsKiii{
D_a^{(1)}={1\over 2\pi i} \pd_{\rho_a} \;\;,\;\;
\tilde D^{(2)}=-{1\over 2 (2\pi i)^2} 
\sum_{a,b} K^{cl}_{ab} \pd_{\rho_a}\pd_{\rho_b}-1 \; .  }
The classical couplings $K^{cl}_{ab}=\int_{X_\D} J_a J_b $ are defined 
for positive bases $J_1, \cdots, J_r $ of $H^2(X_\D,\ZZ)$. 
As a right generalization of the prepotential {\prepotF}
we introduce the prepotential for K3 surfaces by 
\eqn\prepotFiii{
F(t)= \( {1 \over w_0(x)}\)^2 \Big\{ w_0(x) \tilde D^{(2)} w_0(x) 
+ \sum_{a,b}K^{cl}_{ab} D_a^{(1)} w_0(x) D^{(1)}_b w_0(x) \Big\} \;\;,
}
with the mirror map $t_a={D^{(1)}_a w_0(x) \over w_0(x) }$. Then 
the required asymptotics for $F(t)$ will be realized if we 
have the following quadratic relation,  
\eqn\periodrel{
 w_0(x) \tilde D^{(2)} w_0(x) 
+{1 \over2} \sum_{a,b}K^{cl}_{ab} D_a^{(1)} w_0(x) D^{(1)}_b w_0(x) =0 \;\;,  
}
which we may verify explicitly for several examples. 
The fact that we have the required quadratic relation 
for several examples already is convincing for the claim that the our 
arrangement {\wsKiii} of the series provides integral and  
orthogonal basis for the period integrals of the transcendental 
cycles in $H_2(X_\Ds,\ZZ)$.  Note that due to the mirror symmetry 
of the K3 surfaces the symmetric form of 
the transcendental lattice $H_2(X_\Ds,\ZZ)$ is given by   
\eqn\orth{
\Sigma_c=
\( \matrix{ 0 & 0 & 1 \cr 0 & K^{cl}_{ab} & 0 \cr 1 & 0 & 0 \cr } \) \;\;,
}
using the intersection form of the Picard lattice of $X_\D$. 
In the case of  quartic hypersurfaces in $\IP^3$ for example, 
we may verify directly the integrality of the monodromy matrices, as 
well as the orthogonality with respect to {\orth}, through the explicit 
calculations of the monodromy of the hypergeometric series{\wsKiii}. 

In the next sections, we will explain  
these phenomena in the formal expressions {\wsDef}{\wsKiii} 
from the viewpoints of the homological mirror symmetry.

\vfill\eject
\newsec{ Type IIA monodromy }

In this section we write the mirror pair of Calabi-Yau manifolds 
by $X=X_\D$ and $X^\vee=X_\Ds$ for notational simplicity. As will be 
evident, many of the results are not restricted to the hypersurfaces 
in toric varieties.

\subsec{ Homological mirror symmetry and a symplectic basis }

The monodromy of the period integrals are symplectic with respect to 
the intersection form on the cycles of middle dimensions $H_3(X^\vee,\ZZ)$. 
The homology classes of $H_3(X^\vee,\ZZ)$ are considered as D-brane charges 
of (special) Lagrangian 3 cycles in $X^\vee$. Under the homological mirror 
symmetry these cycles are mapped to the elements of $D(X)$, the derived 
category of coherent sheaves. More precisely, Kontsevich proposes categorical 
equivalence between Fukaya's $A_\infty$ category of 
Lagrangian submanifolds 
in $X^\vee$ and the derived category $D(X)$ of the coherent sheaves on 
$X$. Therefore the skew symmetric and integral form on the homology classes 
$H_3(X^\vee,\ZZ)$, as D-brane charges, should have the mirror counterpart. 
The right notion for the D-brane charges is in the K-group $K(X)$. Here we 
consider the topological K-group and the image $K_{hol}(X)$ of $D(X)$ 
assuming elements in $D(X)$ may be represented by suitable complexes of 
vector bundles on $X$ as addressed in section 1. The 
skew symmetric form on $K_{hol}(X)$ may be defined by the alternating sum 
$\chi({\cal E},{\cal F})=
\sum_i (-1)^i {\rm dim} Ext^i_{{\cal O}_X}({\cal E},{\cal F})$ 
for sheaves ${\cal E}$ and ${\cal F}$. If we use 
Riemann-Roch theorem, we may write the skew symmetric form by 
\eqn\RR{
\chi({\cal E},{\cal F})=\int_{X} \ch({\cal E}^\vee \otimes {\cal F}) 
\Todd(X) \;\;.
}
Namely we may realize a skew symmetric form $H^{even}(X,\IQ)$ using 
the Chern character homomorphism $\ch: K(X) \rightarrow H^{even}(X,\IQ)$. We 
should note that the integral structure of the image $\ch(K_{hol}(X))$ 
is different 
from that of $H^{even}(X,\ZZ)$. This is the point we must be careful 
for Calabi-Yau 3-folds. (In case of K3 surfaces the latter is isomorphic to 
the image of $K(X)$.)

With this background in mind, let us introduce a skew symmetric form 
on $H^{even}(X,\IQ)$. First we consider an involution $*$ which acts on 
$H^{2i}(X,\IQ)$ by $(-1)^i$. Using this involution we define 
\eqn\skewEven{
\eqalign{
\langle \alpha,\beta \rangle 
&= \int_X \alpha \wedge * \beta \wedge \Todd(X) \cr
&= - \int_X( \alpha_0 \beta_6 - \alpha_2\beta_4 + \alpha_4\beta_2 
- \alpha_6 \beta_0) \Todd(X) \;,\cr} }
for $\alpha, \beta \in H^{even}(X,\IQ)$.  If we evaluate this skew symmetric 
form for the basis we have introduced in {\sympIIa}, we find that 
it has the `standard' matrix form of the symplectic form, which we 
denote by $\Sigma_c$. Moreover the change of the bases 
\eqn\change{
J_a^{(1)} \rightarrow J_a^{(1)} + \sum_{a,b} C_{ab} J^{(2)}_b \;\;, 
}
with symmetric form $C_{ab}=C_{ba}$ is symplectic, namely does not change 
the symplectic form, although it will change the integral structure 
introduced through the bases in $H^{even}(X,\IQ)$ (in general). 
We should note that 
this change of the bases results in the corresponding change of the asymptotic 
form of the prepotential {\invF} in the quadratic terms of $t$, which are 
known to be non-universal in contrast to other terms. Observing this fact we 
may summarize the results in {\HKTYI}{\HKTYII}{\HLY}{\HLYm} via abundance of 
examples into the following statement;

\vskip0.3cm

\noindent
{\bf Proposition 1}. {\it There exists a canonical symplectic basis 
of the skew symmetric form {\skewEven} on $H^{even}(X,\IQ)$;   
\eqn\sympIIa{
{\bf 1} \;\;,\;\; J_a^{S}
\;\;,\;\;
J_b^{(2)} \;\;, \;\; J^{(3)} \;\;,
}
where $ J_a^{S}=(J_a - \sum_b C_{ab} J^{(2)}_b)(\Todd(X))^{-1}$ 
with some rational constants $C_{ab}=C_{ba}$. 
Corresponding to this basis, we have 
an integral symplectic basis for the period integrals about the LCSL through 
\eqn\IIaIIb{
w_0(x,{J \over 2\pi i}) 
= w_0(x) {\bf 1} + \sum_a D^{(1)}_a w_0(x) J^{S}_a + 
 \sum_{b} \tilde D^{(2)}_b w_0(x) J^{(2)}_b + 
 \tilde D^{(3)}w_0(x) J^{(3)} \;\;, 
}
where $\tilde D^{(2)}_b = D^{(2)}_b + \sum_a C_{ab} D^{(1)}_a$. 
Precisely the hypergeometric series appearing in the coefficients of 
the canonical symplectic basis of $H^{even}(X,\IQ)$ are integral 
symplectic with respect to $\Sigma_c$. 
The corresponding prepotential has the following asymptotic form; 
$F(t) = {1\over3!}\sum_{abc}K^{cl}_{abc} t_a t_b t_c + 
{1\over2}\sum_{ab} C_{ab} t_a t_b - \sum_a {(c_2(X),J_a) \over 24} t_a 
+ {\zeta(3) \over 2 (2\pi i)^3} \chi(X) + {\cal O}(q)$. }

\vskip0.3cm

In the following, we arrange the canonical symplectic basis 
into a row vector 
$ \Pi^{A}(J)=( {\bf 1}, 
J_a^{S}, J^{(2)}_b, J^{(3)} )$ 
and the corresponding period integrals in a column vector $\Pi_{B}(x)$ 
so that we have $w_0(x,{J\over 2\pi i})= \Pi^{A}(J). \Pi_{B}(x)$. 
We should note that the formal relation {\IIaIIb} shows the homological 
mirror symmetry explicitly, as we may read from it the 
correspondences between 3 cycles in the mirror manifold $X^\vee$ and 
the elements in the derived category. The construction of the 
period integrals specifying suitable 3 cycles is one of the hardest part 
in the mirror symmetry.  However our formula {\IIaIIb} replaces this problem 
by easier ones that thinking about coherent sheaves on 
Calabi-Yau manifolds. Now let us study the formula {\IIaIIb} in more details. 

\vskip0.6cm

First of all the hypergeometric series $w_0(x)$ is uniquely characterized 
by the regularity at LCSL. It is known that this series represents the 
period integral for a 3-torus $T^3$ coming from the complex torus $(\IC^*)^4$ 
in the ambient toric variety. This cycle is proved to be special Lagrangian 
for the metric around the degeneration point, the LCSL point. The dual 
cycle to this is a vanishing cycle $S^3$ which appears over the principal 
component of the discriminants of the hypersurface $X^\vee$. The corresponding 
period integral is known to be $\tilde D^{(3)}w_0(x)$ containing the 
highest power of the logarithm. These cycles play an important role for 
the geometric mirror symmetry construction proposed in {\SYZ}. There 
it has been argued in general that the $T^3$ cycle, 
with flat $U(1)$ bundle on it, has its 
deformation space of real dimension 6 whereas the dual cycle $S^3$ with 
flat $U(1)$ bundle is rigid. Under the homological mirror symmetry 
these cycles should be mapped to the sheaves which have the same dimensions 
for their moduli spaces. A little thought tells us that they are given 
by the structure sheaf ${\cal O}_X$ and the skyscraper sheaf ${\cal O}_p$ 
for $S^3$ and $T^3$, respectively.  We evaluate the Chern character 
for each sheaf to be $\ch({\cal O}_X)=1$ and $\ch({\cal O}_p)= \Vol_X$, 
respectively, where $\Vol_X$ is the normalized volume form of $X$. 
Namely we read our formula {\IIaIIb} as  
$$
w_0 {\bf 1} + \cdots + \tilde D^{(3)} w_0 J^{(3)} 
= \ch({\cal O}_X)  \int_{T^3} \Omega  + \cdots + \( - 
\ch({\cal O}_p) \) \int_{S^3} \Omega  \;\;,
$$
where $\Omega$ represents the holomorphic 3-form of the mirror family. 
This relation indicates the following relation should hold in general;
\eqn\homCD{
\langle \ch ({\cal E}_\gamma), w_0(x,{ J\over 2\pi i}) \rangle
= \int_{\gamma} \; \Omega \;\;,
}
where we denote by ${\cal E}_\gamma \in D(X)$ the mirror image of a D-brane 
having the charge $\gamma \in H_3(X^\vee,\ZZ)$. 
As for the from $J_a^S=(J_a-\sum_b C_{ab} J^{(2)}_b )
(\Todd(X))^{-1}$, we may interprete this as the Chern character 
of a torsion sheaf supported on the divisor $J_a$. Suppose for simplicity 
that the divisor $D:=J_a$ is a smooth, irreducible subvariety in $X$. Then 
we may consider a coherent sheaf ${\cal S}$ on the subvariety $D$. We may 
consider this sheaf ${\cal S}$ in the Calabi-Yau manifold extending by zero 
to the outside of $D$. This is the torsion sheaf denoted by 
$i_*{\cal S}$, and whose Chern character may be evaluated 
by Riemann-Roch formula;
\eqn\RRrel{
\ch(i_*{\cal S}) \Todd (X) =i_* \( \ch({\cal S}) \Todd(D) \) \;\;.
}
If we write the components of $\ch(i_*{\cal S})$, we have 
\eqn\RRrelexpand{
\eqalign{
&\ch_1(i_*{\cal S}) = r D \;\;,\cr 
&\ch_2(i_*{\cal S}) = i_*\( c_1({\cal S}) + {r \over 2} c_1(D) \)  \;\;,\cr 
&\ch_3(i_*{\cal S}) = i_*\( ch_2({\cal S})+{1\over2}c_1({\cal S})c_1(D) 
+{r \over 12}\( c_1(D)^2 + c_2(D) \) \) -{r \over 12} D\cdot c_2(X) \;\;, \cr 
}} 
where $r$ is the rank of the sheaf ${\cal S}$ on $D$. Our formula {\sympIIa} 
for $J^S_a$ is essentially this Riemann-Roch formula for the embedding 
of rank one sheaves. The concrete form of the base $J_a^S$ depends on 
the divisors, but we remark here that if a divisor $J_a$  
describes a K3 surface{\KLM}, we may have considerable simplification 
for the choice of the undetermined constants $C_{ab}$. For example, when 
we consider the ideal sheaf ${\cal I}_{p q}$ for the torsion sheaf 
${\cal O}_{pq}$ supported on two points on the K3 surface, we have 
\eqn\idealtwo{
\eqalign{
 \ch(i_*({\cal I}_{pq}))=J_a-{1 \over 12} c_2(X) J_a  \;\;, 
} }
which we evaluate from {\RRrelexpand} 
using $c_1({\cal I}_{pq})=0, \ch_2({\cal I}_{pq})=-2 \Vol_{K3}$ 
as well as $c_1(D)=0, c_2(D)=24 
\Vol_{K3}$.  In the next section we will discuss the cases of del Pezzo 
surfaces after introducing certain automorphisms in the derived category 
$D(X)$, which was proposed by Kontsevich{\KoII} and studied more recently 
by Horja{\Horja}.

\vskip0.5cm

\subsec{ Type IIA monodromy and the monodromy invariant pairing }

Now let us 
define the {type IIA monodromy} (group) acting on the derived category 
$D(X)$ by the mirror transform of the monodromy (group) acting on 
the 3 cycles of the mirror family. 
Then we understand that the actions of the monodromy on the D-brane changes, 
$K_{hol}(X)$, is the same as the corresponding actions on 
$H_3(X^\vee,\ZZ)$ under the mirror equivalence.  
With this definition of the monodromy we may summarize what is 
implied in Proposition 1 as follows;

\vskip0.3cm
\noindent
{\bf ``Theorem'' 2} {\it There exists a monodromy invariant pairing 
\eqn\IinvP{
I: K_{hol}(X) \otimes H_3(X^\vee,\ZZ) \rightarrow  \IQ
}
for the deformation family of Calabi-Yau manifolds $X^\vee$ and its 
mirror manifold $X$. 
}

\vskip0.3cm

If we write the pairing $I({\cal E}_\gamma,\gamma')$ for a symplectic 
basis $\gamma$'s of $H_3(X^\vee,\ZZ)$ and its mirror transform 
${\cal E}_\gamma$'s, 
then our series $w_0(x,{J \over 2\pi i})$ should be understood by
\eqn\womegaI{
w_0(x,{J \over 2\pi i}) = \sum_{\gamma,\gamma'} I({\cal E}_\gamma,\gamma') 
\ch({\cal E}_\gamma) \int_{\gamma'} \Omega \;\;,
}
where the summation is taken over a symplectic basis. Our Proposition 1 
and the formula {\homCD} also say that corresponding to our canonical 
symplectic basis with its symplectic form $\Sigma_c$, the invariant 
pairing has the form  $\( I({\cal E}_\gamma,\gamma') \) = \Sigma_c^{-1}$, 
identifying {\womegaI} with our expansion  
$$
w_0(x,{J \over 2 \pi i})=\Pi^{A}(J). \Pi_{B}(x) =: \Pi_{A}(J) 
\Sigma_c^{-1} \Pi_{B}(x)   \;\;,
$$ 
for $\Pi_A(J)=({\rm ch}({\cal E}_\gamma))$ and 
$\Pi_B(x)=^t(\int_\gamma \Omega)$. Note that our canonical symplectic basis 
$\Pi^A(J)=\Pi_A(J)\Sigma^{-1}_c$ in Proposition 1 is still symplectic because 
$^t\Sigma^{-1}_c = \Sigma_c$.
The monodromy $M$ acting on the period integrals 
$\Pi_{B}(x) \rightarrow \,^t M \Pi_{B}(x)$ should have the type IIA monodromy 
counterpart in the mirror,  $\Pi_{A}(J) \rightarrow \Pi_{A}(J) M$. 
Then our identification of the invariant pairing is consistent because 
we have  the monodromy action on $w_0(x,{J\over 2\pi i})=\Pi_A(J)\Sigma_c^{-1}
\Pi_B(x) \rightarrow \Pi_A(J) M \Sigma_c^{-1} \,^t M\Pi_B(x)=
\Pi_A(J)\Sigma_c^{-1}\Pi_B(x)$ for a symplectic matrix satisfying 
$\,^t M \Sigma_c  M =\Sigma_c$. 

\vskip0.3cm
\noindent
{\bf Remark}. Here we fix our convention about the monodromy actions. 
Let us fix a basis ${\cal S}=\{ \gamma \}$ of $H_3(X,\ZZ)$ and consider 
the Poincar\'e dual $\alpha_\gamma$'s for the basis $H^3(X,\ZZ)$. Then 
we define the skew symmetric form $\Sigma$ by
$$
\int_\gamma \alpha_{\gamma'} = \int_X \alpha_\gamma \wedge \alpha_{\gamma'} 
= \Sigma_{\gamma \gamma'} \;\;.
$$
We consider the monodromy action on the basis $\{ \alpha_\gamma \}$ by 
$\alpha_\gamma \rightarrow \sum_{\gamma'} \alpha_{\gamma'} M_{\gamma'\gamma}$. 
Then the invariance of the skew symmetric form $\Sigma$ is expressed by the 
condition $\,^tM \Sigma M = \Sigma$, i.e., $M$ is symplectic with respect to 
$\Sigma$. We note that by definition of the Poincar\'e dual $\int_\gamma = 
\int_X \alpha_\gamma \wedge$ we have $\int_\gamma \rightarrow \sum_{\gamma'} 
M_{\gamma' \gamma} \int_{\gamma'}$. 
The holomorphic 3-form $\Omega$ is expanded by the basis 
$\{ \alpha_\gamma \}$  as follows;
$$
\Omega=\sum_\gamma {\cal G}_\gamma \alpha_{\gamma} = \sum_{\gamma,\gamma'} 
\alpha_\gamma (\Sigma^{-1})_{\gamma\gamma'} \int_{\gamma'}\Omega \;\;,
$$ 
since we have $\int_\gamma \Omega = 
\sum_{\gamma'}\Sigma_{\gamma\gamma'}{\cal G}_{\gamma'}$. The monodromy 
acts on the period integral ${\cal G}_\gamma$'s by 
${\cal G}_\gamma \rightarrow \sum_{\gamma'} 
(M^{-1})_{\gamma\gamma'}{\cal G}_{\gamma'}$. 
We should compare the above formula 
with our claim {\womegaI} which we reproduce here,
$$
w_0(x,{J \over 2\pi i})=\sum_{\gamma,\gamma'} {\rm ch}({\cal E}_\gamma)
I({\cal E}_\gamma,\gamma') \int_{\gamma'} \Omega \;\;.
$$
Our claim $I({\cal E}_\gamma,\gamma')=(\Sigma^{-1})_{\gamma\gamma'}$ 
is consistent to identify $\Omega$ with  $w_0(x,{J \over 2\pi i})$ under 
the mirror symmetry. If we define the hypergeometric 
series by $\Pi^B(x)=\Sigma_c^{-1} \Pi_B(x)=\,^t( {\cal G}_\gamma )$, then 
the monodromy matrix acts by $\Pi^B(x) \rightarrow M^{-1}\Pi^B(x)$. 
The same arguments applies to Calabi-Yau hypersurfaces (or CICYs)
of arbitrary dimensions. Especially in case of even dimensions, we should 
replace the skew symmetric form $\Sigma$ by corresponding symmetric form, and 
also have the ``period relations'' $\int_X \Omega \wedge \Omega=0$,   
\eqn\relG{
\sum_{\gamma,\gamma'}  
{\cal G}_\gamma \Sigma_{\gamma\gamma'} {\cal G}_{\gamma'} =0 \;\;,}
as a nontrivial consistency condition (, see {\periodrel} for K3 surfaces).

\vskip0.3cm

Now we are ready to present several examples for which integral symplectic 
basis for the period integrals has been obtained explicitly.

\vskip0.5cm

\noindent 
\undertext{{\it Example 1.}}
 (One parameter models in {\CdGP}{\KT}) There are four 
examples of Calabi-Yau hypersurface $X_d (d=5,6,8,10)$ in the weighted 
projective spaces $\IP(\vec \omega)=\IP^4(\omega_1,\cdots,\omega_5)$. 
One is the quintic in $\IP^4$, whose defining data may be written 
$(d;\vec \omega)=(5;1^5)$ specifying the degree and the weights. 
The others have $(d;\vec \omega)=(6;2,1^4),(8;4,1^4),(10;5,2,1^3)$. 
The (even) cohomology ring has a simple form $H^{even}(X,\IQ)=\IQ[J]/(J^4)$ 
with the normalized volume form 
$\Vol_X={{\omega_1\cdots \omega_5}\over d} J^3$. 
The topological data are given 
by $(K_{111}^{cl}, (c_2,J), \chi)=(5,50,-200), (3,42,-204),(2,44,-296),
(1,34,-288)$, respectively for $d=5,6,8,10$. The hypergeometric series 
are determined by the general formula {\fundamentalwo} with 
$l^{(1)}=(-d;\vec \omega)$, and we extract the hypergeometric series from 
the expansion
\eqn\onewoexpand{
w_0(x,{J \over 2\pi i})=w_0(x){\bf 1} + D^{(1)} w_0(x) \; J^S 
+ \tilde D^{(2)} w_0(x) \;  J^{(2)} 
+ \tilde D^{(3)} w_0(x) (-\Vol_X)  \;\;.
}
If we set $J^S = J -{c_2\wedge J \over 12}- a J^{(2)}$ with 
$a={11\over2},{9\over2},3,{1\over2}$, respectively, for $d=5,6,8,10$, 
we may verify that our hypergeometric series $\Pi_B(x)$ reproduces 
integral symplectic basis for period integrals 
constructed explicitly in refs.{\CdGP}{\KT}. (Precisely hypergeometric series 
$^t(-w_0,-D^{(1)}w_0,\tilde D^{(2)}w_0, \tilde D^{(3)}w_0)$ coincides exactly 
with $N \vec \Pi'=N \,^t({\cal G}_1,{\cal G}_2,z^1,z^2)$ in ref.{\KT}, 
see the formula above eq.(5.4) of {\KT} for the matrix $N$. The 
minus signs come from a difference in the overall sign of the prepotential 
${\cal F}$.) See Example 4 for the monodromy calculations. 

\vskip0.5cm
\noindent
\undertext{{\it Example 2.}} ( Two parameter models in {\CdTwoI}{\CdTwoII} ) 
The same calculation proceeds to the two parameter models 
$
X(8) \subset \IP(2,2,2,1,1) \;\;,\;\; X(12) \subset \IP(6,2,2,1,1) 
$  
and $X(18) \subset \IP(9,6,1,1,1)$. For these models integral symplectic 
period integrals for their respective mirror $X(d)^\vee$ 
have been determined in {\CdTwoI}{\CdTwoII}. 
Since the necessary details may be find in the literatures, we simply 
reproduce here the data of their topological couplings; following 
the notation in {\HLY} the non-vanishing classical yukawa 
couplings are summarized in 
$$
 8 J_1^3 + 4 J_1^2 J_2  \;\; {\rm for} \; X(8) \;,\;\;\;
 4 J_1^3 + 2 J_1^2 J_2  \;\; {\rm for} \; X(12) \;,\;\;\;
 9 J_1^3 + 3 J_1^2 J_2 + J_1 J_2^2 \;\; {\rm for} \; X(18) \;,
$$
and also the values of $c_2.\vec J:=((c_2,J_1),(c_2,J_2))$ are 
given by 
$$
(56,24) \;\;  {\rm for} \; X(8) \;\;,\;\;\;\;
(52,24) \;\; {\rm for}  \; X(12) \;\;,\;\;\;\; 
(102,36) \;\; {\rm for} \; X(18) \;\;. 
$$
Note that from these data we may reconstruct 
the ring $H^{even}(X,\IQ)$. For the 
data $\{l^{(1)}, l^{(2)}\}$ we refer to the references. What we need here 
is to find the constants $C_{ab}$ in the basis of 
Proposition 1 which reproduce the results in {\CdTwoI}{\CdTwoII}. 
In ref.{\CdTwoI} these constants are written by 
$\alpha, \beta,\gamma$, related to our constants 
$C_{11}=-\alpha,C_{12}=-\beta, C_{22}=-\gamma $.  For $X(8)$ and $X(12)$, 
it was found that they can be arbitrary integer. Although special values 
$\alpha=0, \beta=-2, \gamma=0$ were chosen there, we set 
$\alpha=\beta=\gamma=0$ in favor of our interpretation {\idealtwo} for the K3 
fibrations. Namely 
the K3 fibrations we see in the divisor $J_2$ simplifies the construction 
of our symplectic basis.  In contrast to these, we need to have 
non-vanishing rational numbers for the $C_{ab}$ of $X(18)$. 
In any case, we can find the constants 
$C_{ab}$ which reproduce the period integrals given in {\CdTwoI}{\CdTwoII}, 
up to the difference above, with the following choices;
\eqn\JsympTwo{
\eqalign{
&X(8): \;\; J_1^S=J_1 - {1\over12}c_2\wedge J_1  \quad,\quad 
            J_2^S=J_2 - {1\over12}c_2\wedge J_2  \cr
&X(12): \; J_1^S=J_1 - {1\over12}c_2\wedge J_1   \quad,\quad 
           J_2^S=J_2 - {1\over12}c_2\wedge J_2  \cr
&X(18): \; J_1^S=J_1 - {1\over12}c_2\wedge J_1- {9\over 2} J_1^{(2)}
                                          - {3\over2}J_2^{(2)}   \;,\; 
           J_2^S=J_2 - {1\over12}c_2\wedge J_2 - {3\over2}J_2^{(2)}.  \cr
} }
For $X(18)$, it may be convenient to define another basis 
$H=J_1, E=J_1-3 J_2$ 
for $H^{even}(X,\IQ)$ since in this basis the classical yukawa couplings 
simplifies to $9 H^3 + 9 E^3$. This is because the divisor $E$ is 
isomorphic to $\IP^2$ which may be contracted to a point(, 
see the next section). 
%

\vskip0.3cm

\noindent 
{\bf Remark.} In case of the lattice polarized K3 surfaces, there is no 
ambiguity in our expansion $w_0(x,{J \over 2\pi i})$ presented in {\wsKiii}. 
In this case the identifications of our bases {\KiiiBasis} with suitable 
coherent sheaves on $X$ are 
\eqn\identSheaves{
{\bf 1}' = \ch({\cal I}_p) \;\;,\;\; J^{(2)}=-\Vol_X = -\ch({\cal O}_p) \;\;,
}
and $J_a=\ch({\cal O}_{D_a}(L))$ with the torsion sheaf supported on the 
divisor $D_a=J_a$ tensored with a suitable line bundle $L$ on it. 
The line bundle $L$ on $D_a$ is determined requiring $\ch({\cal O}_{D_a}(L))
=D_a-{D_a^2 \over 2} + ({\rm deg}L) \, \Vol_{K3}=D_a$. 
Note that the Chern character of the 
ideal sheaf ${\cal I}_p$ for the skyscraper sheaf ${\cal O}_p$ may be 
evaluated from the exact sequence $0 \rightarrow {\cal I}_p \rightarrow 
{\cal O}_X \rightarrow {\cal O}_p \rightarrow 0$. Note also that 
the sheaf ${\cal I}_p$ is the image of ${\cal O}_p$ under the Fourier-
Mukai transform $\Phi_{{\cal Q}^\bullet}(\cdot): D(X) \rightarrow D(X)$ 
defined by the kernel ${\cal Q}$, the ideal sheaf 
$0\rightarrow {\cal Q} \rightarrow {\cal O}_{X\times X} \rightarrow 
{\cal O}_\D \rightarrow 0$ for the diagonal $\Delta \subset X\times X$ 
{\MukaiII}(cf. definitions in the next section). 

The monodromy invariant pairing {\IinvP} in this case should be understood 
$I: K_{hol}(X) \otimes H_2(X^\vee,\ZZ)_{trans} \rightarrow \IQ$, where 
$H_2(X^\vee,\ZZ)_{trans}$ is the transcendental lattice. For a basis of 
the transcendental lattice whose  orthogonal form $\Sigma_c$ given in 
{\orth}, we have $\(I({\cal E}_\gamma,\gamma)\)=\Sigma_c^{-1}$ and thus 
$w_0(x,{J \over 2\pi i})=\sum_{\gamma,\gamma'} \ch({\cal E}_\gamma)
({\Sigma_c^{-1}})_{\gamma,\gamma'} \int_{\gamma'} \Omega$ where 
$\Omega$ represents the holomorphic 2 form. Full determination 
of the type IIA monodromy group is beyond the scope of this paper, however 
we may expect that the monodromy group would be generated by the 
$(-2)$-reflections due to Mukai on $D(X)${\MukaiII}.

The case of the elliptic curves is more simple and we may 
identify the type IIA monodromy with $SL(2,\ZZ)$ action on $D(X)$ 
in {\MukaiI}. This is clear but we will evaluate explicitly the 
monodromy of the relevant hypergeometric series in section 3.4.

\subsec{ Automorphisms in derived categories }

In 1996 Kontsevich{\KoII} considered certain automorphisms of $D(X)$ 
which come 
naturally from the homological mirror symmetry, namely Fourier-Mukai 
transforms with suitable choices of its kernels. To describe them let 
us consider for a Calabi-Yau 3 fold $X$ the diagram: 
\eqn\diag{
\matrix{ X \times X  & \mapright{p_2} & X \cr
         \mapdown{p_1} &  & \cr
         X &  &  \cr }
}
When we fix an object ${\cal E}\in D(X\times X)$, we may consider a 
functor $\Phi_{{\cal E}^\bullet}(\cdot)=\IR^{\bullet} {p_2}_*({\cal E}
{\otimes^{\bf L} } p_1^* (\cdot)): D(X) \rightarrow D(X)$. If the functors 
of this type define equivalences (automorphisms) of the category, they 
are called Fourier-Mukai transforms. We call the object ${\cal E} \in 
D(X \times X)$ in $\Phi_{\cal E}$ the {\it kernel} of the functor.  
The conditions on the kernel for $\Phi_{\cal E}$ to define 
a Furier-Mukai transform has been  studied in general for smooth projective 
varieties in {\Bri}{\BO}. The Fourier-Mukai transform $\Phi_{\cal E}: 
D(X) \rightarrow D(X)$ induces corresponding automorphism 
$K_{hol}(X) \rightarrow K_{hol}(X)$ under the canonical map, 
and futher an automorphism 
on $H^{even}(X,\IQ)$ using the Chern character ring homomorphism $\ch: K(X) 
\rightarrow H^{even}(X,\IQ)$. 
The identity may be realized by the torision sheaf 
${\cal O}_\D \in D(X\times X) $ supported on the diagonal 
$\D \subset X\times X$, namely the delta function supported on the diagonal.

Now let us introduce the kernels proposed by Kontsevich. The first kernel 
is given by the complex,
\eqn\lineL{
\cdots \rightarrow 0 \rightarrow {\cal O}_\D \otimes p_2^*({\cal L}) 
\rightarrow 0 \rightarrow \cdots  \;,
}
where ${\cal L}$ is the sheaf of a line bundle on $X$ and the non-zero 
sheaf is at the zero-th position. The corresponding Furier-Mukai transform 
acts on a sheaf ${\cal F} \in D(X)$ as ${\cal F} \mapsto 
{\cal F}\otimes {\cal L}$. Since in our case of hypersurfaces in toric 
varieties  the Picard group are generated by our basis $J_1, \cdots , J_r$ of 
$H^{1,1}(X,\ZZ)$, we may consider $r$-independent kernels of this type. 
Taking the Chern character, we may summarize the Furier-Mukai transforms 
of this type by
\eqn\gammaE{
{\cal D}_{x_k}: \gamma \mapsto {\rm e}^{J_k} \gamma \;\;\quad
\quad (\gamma \in H^{even}(X,\IQ); \; k=1,\cdots,r ). 
}
We have used the notation ${\cal D}_{x_k}$ because we can identify these 
automorphisms with the type IIA monodromy corresponding to the monodromy 
about the toric divisors $x_k=0$ at the LCSL. The second is the 
kernel whose cohomology is the ideal sheaf ${\cal I}_\D$ of the diagonal;
\eqn\diszero{
\cdots \rightarrow 0 \rightarrow 
{\cal O}_{X\times X} \rightarrow {\cal O}_\D \rightarrow 0 
\rightarrow \cdots , }
where ${\cal O}_\D$ at the zero-th position. The functor 
$\Phi_{{\cal E}}$ acting on a sheaf ${\cal F}$ results in $\cdots \rightarrow 
0 \rightarrow {p_2}_*({\cal O}_{X \times X} \otimes p_1^* {\cal F}) 
\rightarrow {\cal F} \rightarrow 0 \rightarrow \cdots$, which implies the 
following automorphism on $H^{even}(X,\IQ)$,
\eqn\Tzero{
{\cal T}_0 : \gamma \mapsto \gamma 
- \(\int_{X} \Todd(X)\wedge\gamma \) {\bf 1}  \;\;\quad (\gamma 
\in H^{even}(X,\IQ) ).
}
Kontsevich proposed that this should reproduce the monodromy about the 
principal components of the discriminant of the mirror family $X^\vee$. 
Since over the principal discriminant we have vanishing cycle(s), $S^3$, 
we may identify this monodromy with the Picard-Lefschetz formula. 

\vskip0.6cm

Recently Horja{\Horja}, generalized the kernel {\diszero} to more general 
situation, which we may understand the local mirror symmetry counterpart 
of the formula {\Tzero}. 
Here we briefly summarize his result to clarify its relation to the 
local mirror symmetry. Let us assume an exceptional locus $E$ in 
a Calabi-Yau hypersurface (or CICY) $X$ which may 
be contracted to a locus $Z$ in $X$.  In the ambient toric varieties 
the exceptional locus $E_0$ of the (elementary) contractions are given 
by the intersections of toric divisors, $D_{i_1}\cap\cdots\cap D_{i_k}$. 
Here we assume our exceptional 
locus $E$ comes from the ambient space, namely $E_0$ restricted to $X$. 
This is exactly the local mirror symmetry considered in {\KMV}{\CKYZ}. 
In this setting we may consider the following diagram,
\eqn\diagLocal{
\matrix{ E \times_Z E  & \mapright{r_2} & E \cr
         \mapdown{r_1} &  & \cr
         E &  &  \cr }
}
together with the embeddings $j_a: E \hookrightarrow X \; (a=1,2)$ and $j:  
E \times_Z E \hookrightarrow X \times X$. We write $Y:=E\times_Z E$. 
Given an object ${\cal I} \in D(Y)$ we may consider the functor 
$\Phi'_{{\cal I}^\bullet}(\cdot)=\IR^\bullet (j_2\circ r_2)_* ( {\cal I} 
\otimes^{\bf L} (j_1\circ r_1)^*(\cdot)): D(X) \rightarrow D(X)$. Since we 
may verify $\Phi'_{{\cal I}}(\cdot)=\Phi_{j_*{\cal I}}$ 
(Lemma 4.4 of {\Horja}), we obtain 
new kernels of the Fourier-Mukai transforms in {\diag} in this way. 
For the local mirror symmetry version of Picard-Lefschetz formula {\Tzero}, 
a naive choice of the kernel would be to consider the complex ${\cal I}:$
$\cdots \rightarrow 0 \rightarrow {\cal O}_Y \rightarrow 
{\cal O}_{\D'} \rightarrow 0 \rightarrow \cdots$, where $\D' \subset 
Y=E\times_Z E$ is the diagonal ($\D'\cong E$). The actual construction 
is more involved using the (reduced) Koszul complex for ${\cal O}_{\D'}
\cong {\cal O}_E$ and the mapping cone construction, (see ref.{\Horja} for 
full details). The corresponding automorphisms on $H^{even}(X,\IQ)$ are 
determined (Proposition 4.5 {\Horja}) to be
\eqn\horjaI{
\gamma \mapsto \gamma - \prod_{a=1}^k (1-{\rm e}^{D_{i_a}}) 
{r_2}_*\( (j_1\circ r_1)^*(\gamma)\, \Todd(Y)\, r_2^*(\Todd(E)^{-1}) \) \;,
}
for $\gamma \in H^{even}(X,\IQ)$.

In our case of toric varieties, hypersurfaces (of CICYs) in toric 
varieties, the contraction shows itself 
as an extremal ray of the Mori cone (or its higher dimensional cone). 
In this paper we restrict our attention to the primitive contractions, 
where we have an extremal ray representing the contraction. Let us 
write the extremal ray $l^{(k)}=(l_0^{(k)}; l_1^{(k)}, \cdots, l_p^{(k)})$ 
in our notation prepared in section 2.1. Geometrically extremal ray 
represents a rational curve collapsed under the contraction. 
The duality of the lattice $L$ to $H^{1,1}(X,\ZZ)$ (, precisely 
$A^1(\IP_\SDs)$) comes from the intersection pairing between algebraic 
2 cycles and the divisors. From this reason we may read off the 
divisor which contains 
the curves to be contracted from the extremal ray, namely, we simply read the 
negative entry of the vector $l^{(k)}$. Based on these facts we make 
the following definitions,  
$I_{+}=\{ i \;|\; l_i^{(k)} > 0 \; \}, \; 
I_-'=\{ i \;|\; l_i^{(k)} <0 (1\leq i\leq p)\;\} $, 
$I_-''=\{ 0\}$ if $l_0^{(k)} <0$, 
and $I_-''=\{\phi\}$ otherwiae, 
where $I_-'$ represents the toric 
divisors to be contracted (, the case $l_0^{(k)}>0$ does not occur 
for $\IP_\SDs$ being projective). We may also read the dimensions of $Z$ 
from the cardinality of the set $I_{+}$. 
Corresponding to these primitive contractions characterized by $l^{(k)}$, 
the automorphisms {\horjaI} are written more explicitly (Proposition 4.20 
{\Horja});
\eqn\horja{
{\cal T}_k : \gamma \mapsto \gamma - \prod_{i\in I_{-}'}(1-{\rm e}^{D_i})
\int_{C_\xi} { \prod_{i \in I_{-}''} (1-{\rm e}^{D_i(\xi)}) \over 
                \prod_{i \in I_{+} } (1-{\rm e}^{-D_i(\xi)}) } \gamma(\xi) 
{d \xi \over 2\pi i} \;\;,
}
where $D_i$ represents the toric 
divisors represented by the vertices of $\Ds$. Note that we may write 
$D_i=\sum_{m} l_i^{(m)} J_m$ because $J_1, \cdots , J_r$ 
are defined to be the dual basis to $l^{(1)},\cdots, l^{(r)}$. Then 
$D_i(\xi)$ in the above formula is defined by 
$D_i(\xi):= l_i^{(k)} \xi + \sum_{m\not=k} l_i^{(m)} J_m$ 
with a formal variable $\xi$. $\gamma(\xi)$ is defined by the 
corresponding product of $D_i(\xi)$'s to the representation 
$\gamma=\prod D_i$. 
The contour integral are understood to take all the residues about 
the zeros of the denominator. 

\vskip0.6cm

In our compactification based on the secondary fan, which is briefly 
summarized in section 2.1, the Mori cone has a meaning to describe the 
coordinate ring of an affine chart of the compactified moduli space 
for the mirror family of $X^\vee$. 
In this toric compactification, the extremal ray (one dimensional cone) 
describes coordinate ring for one dimensional torus invariant subvariety 
isomorphic to $\IP^1$. This is exactly the $\IP^1$ whose coordinate is 
given by $x_k$ and other variables vanish. It has been proved 
over $\IC$ {\Horja} that the automorphism above 
coincides with the monodromy of the GKZ system 
around a certain loop contained in the $\IP^1$. 
What we have now is a natural integral structure 
which comes from our canonical integral symplectic basis {\sympIIa} 
of $H^{even}(X,\IQ)$. In fact we may verify, up to conjugation, 
that the automorphisms 
${\cal D}_{x_k}$, ${\cal T}_0, {\cal T}_k$ on $H^{even}(X,\IQ)$ 
reproduce exactly the monodromy determined in {\KT}{\CdGP} as expected 
from our observation summarized in Example 1 and Example 2. Furthermore 
we may verify for many examples that these automorphisms are integral 
and symplectic with respect to our symplectic basis in Proposition 1 
if we take a suitable choice for the undetermined rational constants $C_{ab}$.

\vskip0.5cm

\subsec{ Monodromy of the elliptic curves and some other examples}

As the simplest cases of the type IIA monodromy, we present examples of 
the elliptic curves; $E_6:$ the cubic in $\IP^2$, $E_7:$ the degree four 
curves in $\IP^2(2,1,1)$ and $E_8:$ the degree six curves in $\IP^2(3,2,1)$. 
For the degree $d$ curve in $\IP^2(a,b,c)$ the toric data may 
be read from the dual 
polytope $\Ds(\vec \omega)={\rm Conv}.\( (1,0),(0,1),(-a,-b) \)$. For example, 
we determine the basis of the Mori cone to be
$l=(-d; a,b,c) \;\;,\;\; (d=a+b+c)$, and also $H^{even}(X,\IQ)=\IQ[J]/(J^2)$ 
with the normalizaton $\int_X J = 1, 2, 3$, respectively for 
$X=E_8,E_7$ and $E_6$. Given these data we may write the hypergeometric 
series;
$$
w_0(x, { J \over 2\pi i}) = w_0(x) + J {1\over 2\pi i} \pd_\rho w_0(x) 
\;\;.
$$  
Since the prepotential of the elliptic curves is given simply by $F(t)=t$, 
there in no amibiguity to define the canomical ``symplectic'' basis 
of $H^{even}(X)$. It is fixed by $\Pi^{A}(J)=({\bf 1}, -{1\over d'}J)$, 
where $d'={d \over a b c}$, and 
correspondingly integral symplectic basis for the cycles should be given  
by $\Pi_{B}(x)=\;^t(w_0, - {d' \over 2\pi i} \pd_{\rho} w_0 )$ accoding to 
our pairing. The analytic 
continuations of the hypergeometric series are simple exercises, for example, 
we may follow the calculation in {\CdGP} based on the Barnes integral 
representations. In Table 1, 
we list the connection matrix $N$ of the series about $x=0$ to those about 
$x=\infty$, from which we may determine the momodromy about $x=0$ and the 
discriminant $dis_0(x)=1-27 x, 1-64 x, 1-432 x$, repectively for 
$X=E_6,E_7, E_8$. For readers convenience we present the hypergeometric series 
about $x=\infty$;
$$
\eqalign{
& w^\infty_j(x)=\sum_{k=1}^{d-1} (1-\alpha^{a k})(1-\alpha^{b k})\alpha^{kj}
\tilde w_k(x)   \;, \cr
& \tilde w_k(x)=-{1\over d}{1\over (2\pi i)^2} \sum_{N\geq 0} 
{\Gamma(a N+{a k \over d})\Gamma(b N + {b k \over d})\Gamma(N+{k\over d}) 
\over \Gamma(d N+k) } x^{-N-k/6} \;, \cr}
$$
where $\alpha$ is the primitive root of $\alpha^d=1$ for $d=a+b+c$.  

$$
\def\vspace#1{ \omit & \omit & height #1 & \omit && \omit && \omit &\cr }
\vbox{\offinterlineskip
\hrule
\halign{ \strut
 \vrule#&  $\;$ \hfil #  \hfil  
&\vrule#&  $\;$ \hfil # \hfil 
&\vrule#&  $\;$ \hfil # \hfil  
&\vrule#&  $\;$ \hfil # \hfil  
&\vrule# 
\cr 
&   &&  $E_6$  &&  $E_7$  &&  $E_8$  &\cr
\noalign{\hrule} 
\vspace{1pt}
\noalign{\hrule} 
& $N$    && $\( \matrix{ 1 & 0 \cr {1\over3} & -{1\over3}\cr} \)$
         && $\( \matrix{ 1 & 0 \cr {1\over2} & -{1\over2}\cr} \)$
         && $\( \matrix{ 1 & 0 \cr {1} & -{1}\cr} \)$
          &\cr
\noalign{\hrule} 
& $M_0$  &&  $\( \matrix{ 1 & 0 \cr -3 & 1 \cr} \)$
         &&  $\( \matrix{ 1 & 0 \cr -2 & 1 \cr} \)$
         &&  $\( \matrix{ 1 & 0 \cr -1 & 1 \cr} \)$
         &\cr
& $M_1$  &&  $\( \matrix{ 1 & 1 \cr 0 & 1 \cr} \)$
         &&  $\( \matrix{ 1 & 1 \cr 0 & 1 \cr} \)$
         &&  $\( \matrix{ 1 & 1 \cr 0 & 1 \cr} \)$
         &\cr
}
\hrule}
$$
{\leftskip1cm\rightskip1cm\noindent
{\bf Table 1}.  The connection matrix $N$ and the monodromy matrices 
about $x=0$ and the discriminant $dis_0(x)$. 
The matrices $M_0$ and $M_1$ generate the modular group $\Gamma_0(3), 
\Gamma_0(2)$ and the modular group $\Gamma=SL(2,\ZZ)$ for $E_6,E_7, E_8$, 
respectively. The connection matrix relates the hypergeometric 
series by $\;^t( w_0, -{d'\over 2\pi i}\pd_{\rho} w_0 ) = 
N \;^t( w^\infty_0,  w^\infty_1 )$.  \par}

\vskip0.3cm

As we see in the table, $\Pi_{B}(x)$ in fact provides an integral, 
symplectic basis for the period integrals. Furthermore it is easy to verify 
that the type IIA monodromy ${\cal T}_0$ and ${\cal D}_x$ exactly reproduce 
the the corresponding monodromy matrices when acted on the ``symplectic'' 
basis $\Pi^{A}(J)$. The modular group $\Gamma=SL(2,\ZZ)$ is nothing but 
the modular group realized in the derived category $D(X)$ by Mukai{\MukaiI}. 

\vskip0.3cm

We can extend these analysis for the lattice polarized 
K3 surfaces observing that there is no ambiguity in defining our 
canonical ``orthogonal'' basis for the Picard lattice. Our ``Theorem'' 2 
(or more concretely the equation {\wsKiii}) implies that we will obtain 
an orthogonal basis for the period integrals in this way.  

In the rest of this section we present explicit examples of the monodromy 
calculations for simple cases. 

\vskip0.5cm

\noindent
\undertext{{\it Example 3.}} (Quartic hypersurface in $\IP^3$) 
The quartic hypersurface $X(4)$ in the projective space $\IP^3$ has the 
Picard lattice $\langle 4 \rangle$, i.e., the rank one lattice 
with its generator $J$ of the intersection number $4$. This is a simple 
example of the lattice polarized K3 surface, and has its toric data 
$\D(1^4)^*={\rm Conv.}\( (1,0,0),(0,1,0),(0,0,1),(-1,-1,-1) \)$. In Batyrev's 
mirror construction we may write $X(4)=X_{\D}$ using the dual polytope $\D=
\D(1^4)$. The cohomology ring $H^{even}(X(4),\ZZ)$ is described explicitly by 
$\ZZ[J]/(J^3)$. The hypergeometric series representing the period integral 
of the mirror $X(4)^\vee=X_{\Ds}$ may be determined by the formula 
{\fundamentalwo} with $l=(-4;1,1,1,1)$. Then following {\wsKiii} we have
\eqn\quarticWo{
w_0(x,{J\over 2\pi i}) = 
w_0(x) \( 1- {J^2 \over 4}\) + D^{(1)}w_0(x) \, J + 
D^{(2)}w_0(x) \( - {J^2 \over 4} \) \;\;.
} 
We identify this expansion by $w_0(x,{J \over 2\pi i})=
\Pi_{A}(J).\Pi^{B}(x)$ introducing the notations $\Pi_{A}(J)=
(1-{J^2 \over 4}, J, -{J^2 \over 4} )$, and $\Pi^{B}(x)=\,^t(w_0(x), 
D^{(1)}w_0(x), \tilde D^{(2)}w_0(x) )$. The way of identification is slightly 
differs from that in Proposition 1 for 3 folds cases, since we verify 
right intersection form for the basis $\Pi_A(J)$ (, see below). 
More explicitly we have 
\eqn\Dwos{
w_0(x) \;\;,\;\;
D^{(1)}w_0(x)= \pd_\rho w_0(x) \;\;,\;\;
\tilde D^{(2)}w_0(x)=-2 \pd_\rho^2 w_0(x) - w_0(x) \;\;,
}
for the period integral about $x=0$. The analytic continuation to 
$x=\infty$ is similar to the cases in elliptic curves. Since it is 
easy to determine the monodromy about $x=0$ and $x=\infty$, we can 
determine the monodromy group once we know the connection formula 
of the two local solutions. In Appendix, we briefly summarize 
the monodromy calculation 
done in {\CdGP} applying to the present case. (The calculations are 
essentially the same for all monodromy calculations done in this 
paper.) In any case we obtain the monodromy matrices $M_0$, $M_\infty$ and,  
in turn, $M_0$ and $M_1=(M_0 M_\infty)^{-1}$ after determining the 
connection matrix $N$ (, where $\Pi^{B}(x)=N \Pi^{B}_\infty(x)$ with 
$\Pi^B_\infty(x)=\,^t(w_0^\infty, w_1^\infty, w_2^\infty)$, see Appendix for 
the definition $w_j^\infty(x)$,);
\eqn\Mquartic{
M_0=\( \matrix{ 1 & 0 & 0 \cr 1 & 1 & 0 \cr -2 & -4 & 1 \cr} \) \;,\; 
M_1=\( \matrix{ 0 & 0 & 1 \cr 0 & 1 & 0 \cr  1 &  0 & 0 \cr} \) \;,\; 
N=\( \matrix{ 1 & 0 & 0 \cr -{1\over4} & {1\over2} & {1\over4} \cr 
             -{1\over2} & -{1\over2} & 0 \cr} \) \;.\; 
}
Now it is an easy exercise to verify the same monodromy matrices $M_0$ and 
$M_1$ from the Fourier-Mukai transformations {\gammaE} and {\Tzero}, 
respectively, in the basis $\Pi_{A}(J)$. (Precisely we should have 
$\Pi_{A}(J) \rightarrow \Pi_{A}(J) M^{-1}$ if $\Pi^B(x) \rightarrow 
M \Pi^B(x)$ according to Remark after {\womegaI}. 
However this is simply a matter of the identification 
of the monodromy actions, i.e., either we reverse the monodromy actions 
on $\Pi^{B}(x)$ or reverse the Fourier-Mukai transformations {\gammaE}, 
{\Tzero}.) 
%
%
Also we may verify that the monodromy matrices $M_0$ and 
$M_1$ are orthogonal with respect to the canonical orthogonal form 
$\Sigma_c$ in {\orth}. We should note that the orthogonal form {\orth} 
is nothing but the symmetric form $\langle \alpha, \beta \rangle 
= - \int_X(\alpha_0\beta_4- \alpha_2\beta_2)\Todd(X)$ {\skewEven} expressed 
in the basis $\{ 1-{J^2 \over 4}, J, -{J^2 \over 4} \}$. Corresponding 
period integrals $\Pi^{B}(x)$ are identified with the period integrals 
of the transcendental cycles of the mirror $X(4)^\vee$. As a consistency 
we verify explicitly the period relation {\relG}, or explicitly using 
{\periodrel},
\eqn\quartRel{
w_0(x) D^{(2)}w_0(x)+{4 \over 2}\( D^{(1)}w_0(x) \)^2 =0 \;\;. 
}
The quadratic relation above may be used to determine the modular 
property of the mirror map $x=x(q)$ defined by 
$t={D^{(1)}w_0(x) \over w_0(x)}$ with $q={\rm e}^{2\pi i t}$. Namely 
the monodromy $M_0$ simply acts on $t$ by $t \rightarrow t+1$, while the 
monodromy $M_1$ acts
$$
t={D^{(1)}w_0(x) \over w_0(x)} 
\rightarrow 
{D^{(1)}w_0(x) \over D^{(2)}w_0(x)}=
{w_0(x) D^{(1)}w_0(x) \over w_0(x) D^{(2)}w_0(x)}
=-{1 \over 2t} \;\;.
$$
Now we recall that the modular group $\Gamma_0(N)_+$ is a normalizer of 
$\Gamma_0(N)$ in $SL(2,\IR)$, defined by adding the Fricke involution 
$t \rightarrow -{1 \over N t}$. Since for $N=2$ we may write 
$\Gamma_0(2)+=\langle \(\matrix{0 & -1 \cr 2 & 0 \cr}\), 
\(\matrix{1 & 1 \cr 0 & 1 \cr}\) \rangle$, we identify the modular group 
of our mirror map $x=x(q)$ with $\Gamma_0(2)_+$. In this way we may arrive 
at the known result in {\LY}.

\vskip0.5cm

\noindent
\undertext{{\it Example 4.}} (One parameter models in {\CdGP}{\KT}) 
For readers convenience we list the monodromy matrices for the 
one parameter models in Example 1. Although the results are obtained in 
{\CdGP}{\KT}, we reproduce them here for the comparison with the type IIA 
monodromy calculations. The monodromy calculations for the hypergeometric 
series $\Pi_B(x)=\,^t( w_0, D^{(1)}w_0, \tilde D^{(2)}w_0, 
\tilde D^{(3)} w_0 )$ are exactly the same as in Example 3. 
We define the hypergeometric series about $x=\infty$ by 
\eqn\infsolone{
\eqalign{
w_j^\infty(x)&=\sum_{k=1}^d 
(\prod_{i=1}^4 (1-\alpha^{\omega_i k}))\alpha^{kj} \tilde w_k(x) 
\;\; (j=0,1,\cdots,d-1) \;,\cr
\tilde w_k(x)&=-{1\over d}{1\over (2\pi i)^4}
\sum_{N=0}^\infty { \prod_{i=1}^5 \Gamma(\omega_i(N+{k\over d})) \over \Gamma(d N+k) } x^{-N-{k\over d}} \;. 
} }
All of these series are not linearly independent but have some relations, 
for example ,$w_0^\infty + \cdots + w_{d-1}^\infty=0$ (, see {\KT}). 
Taking an independent set we define 
$\Pi_B^\infty(x)=\,^t( w_0^\infty(x), w_1^\infty(x),$ $ 
w_2^\infty(x), w_{d-1}^\infty(x))$ and the connection matrix $N$ by the 
relation $\Pi_B(x)=N \Pi_B^\infty(x)$. Since the monodromy about $x=\infty$ 
is simple in the basis $\Pi_B^\infty(x)$, we can determine the corresponding 
action on $\Pi_B(x)$ using the connection matrix. The results are listed 
in Table 2. 

We leave for reader's exercise to verify the same monodromy matrices 
follow from the type IIA monodromy, namely 
Fourier-Mukai transformations {\gammaE}{\Tzero} acting on our canonical 
symplectic basis 
$\Pi^A(J)=(1, J-{c_2\wedge J \over 12}-a J^{(2)}, J^{(2)}, -{\rm Vol}_X)$ 
with $a={11\over2},{9\over2},3,{1\over2}$, respectively for $d=5,6,8,10$ (, 
see Example 1).  The action {\Tzero} reproduces the monodromy about 
the discriminant, which is in the form of Picard-Lefshetz fromula for 
all $X_d$; 
$$
M_1=(M_0 M_\infty)^{-1}=\(\matrix{1 & 0 & 0 & 1 \cr 0 & 1 & 0 & 0 \cr 
0 & 0 & 1 & 0 \cr 0 & 0 & 0 & 1 \cr} \) \;\;.
$$

$$
\vbox{ \tabskip=0pt
\offinterlineskip
\halign{ 
\strut# &\vrule# 
&\hfil$\;\;#\;\;$\hfil  
&\vrule#
& \hfil$\quad#\quad$\hfil 
&\vrule#
& \hfil$\quad#\quad$\hfil 
&\vrule#
& \hfil$\quad#\quad$\hfil 
& \vrule#  \cr
\noalign{\hrule}
\omit && d && M_0 && M_\infty  && N  &\cr
\noalign{\hrule}
\omit&& 5 
&& \( \matrix{ 1 & 0 & 0 & 0 \cr 1 & 1 & 0 & 0 \cr 8 & 5 & 1 & 0 \cr -5 & 3 & 
     -1 & 1 \cr  } \)
&&  \( \matrix{ 1 & 0 & 0 & -1 \cr -1 & 1 & 0 & 1 \cr -3 & -5 & 1 & 3 \cr 5 & 
     -8 & 1 & -4 \cr  } \)
&& \( \matrix{ 1 & 0 & 0 & 0 \cr -{\frac{2}{5}} & {\frac{2}{5}} & {\frac{1}
      {5}} & -{\frac{1}{5}} \cr -{\frac{21}{5}} & {\frac{1}{5}} & {\frac{3}
     {5}} & -{\frac{8}{5}} \cr 1 & -1 & 0 & 0 \cr  } \)
&\cr
\noalign{\hrule}
\omit&& 6 
&& \( \matrix{ 1 & 0 & 0 & 0 \cr 1 & 1 & 0 & 0 \cr 6 & 3 & 1 & 0 \cr -4 & 3 & 
     -1 & 1 \cr  } \)
&& \(\matrix{ 1 & 0 & 0 & -1 \cr -1 & 1 & 0 & 1 \cr -3 & -3 & 1 & 3 \cr 4 & 
     -6 & 1 & -3 \cr  } \)
&& \( \matrix{ 1 & 0 & 0 & 0 \cr -{\frac{1}{3}} & {\frac{1}{3}} & 
   {\frac{1}{3}} & -{\frac{1}{3}} \cr -3 & 0 & 1 & -2 \cr 1 & -1 & 0 & 0 
   \cr  } \)
&\cr
\noalign{\hrule}
\omit&& 8
&& \( \matrix{ 1 & 0 & 0 & 0 \cr 1 & 1 & 0 & 0 \cr 4 & 2 & 1 & 0 \cr -4 & 2 & 
     -1 & 1 \cr  } \)
&& \( \matrix{ 1 & 0 & 0 & -1 \cr -1 & 1 & 0 & 1 \cr -2 & -2 & 1 & 2 \cr 4 & 
     -4 & 1 & -3 \cr  } \)
&& \( \matrix{ 1 & 0 & 0 & 0 \cr -{\frac{1}{2}} & {\frac{1}{2}} & {\frac{1}
     {2}} & -{\frac{1}{2}} \cr -3 & 0 & 1 & -2 \cr 1 & -1 & 0 & 0 \cr  } \)
&\cr
\noalign{\hrule}
\omit&& 10
&& \(\matrix{ 1 & 0 & 0 & 0 \cr 1 & 1 & 0 & 0 \cr 1 & 1 & 1 & 0 \cr -3 & 0 & 
     -1 & 1 \cr  } \)
&& \( \matrix{ 1 & 0 & 0 & -1 \cr -1 & 1 & 0 & 1 \cr 0 & -1 & 1 & 0 \cr 3 & 
     -1 & 1 & -2 \cr  } \)
&& \( \matrix{ 1 & 0 & 0 & 0 \cr 0 & 0 & 1 & -1 \cr -1 & -1 & 0 & -1 \cr 1 & 
     -1 & 0 & 0 \cr  } \)
&\cr
\noalign{\hrule} } }
$$
{\leftskip1cm \rightskip1cm \noindent
{\bf Table 2}.  The connection matrix $N$ and the monodromy matrices for 
one parameter models $X_d$ $(d=5,6,8,10)$. Matrices are written in terms 
of our canonical symplectic basis defiend by $w_0(x,{J \over 2\pi i})=
\Pi^A(J).\Pi_B(x)$.  \par}


\vskip1cm

\newsec{ Local limit of the monodromy }

\subsec{ Calabi-Yau models }

Here we briefly summarize some properties of elliptic Calabi-Yau 3-folds 
over the Hirzebruch surface ${\bf F}_1$.  
These Calabi-Yau 3-folds are studied extensively in the context of 
F-theory{\MVI}{\MVII}{\KMV}. We will study in detail the local 
mirror symmetry limit of their monodromy. 

Let us consider the affine space 
$\IC^7$ with its coordinate $x_1,x_2,\cdots,x_7$ and the torus actions 
generated by 
\eqn\Cabc{
({\bf C}^*)_{(a,b,c)}^3= 
 \Bigg\langle \matrix{ (a,b,\;\,c,0,0,0,\;\,0)_\lambda, \cr 
                  (0,0,-1,1,1,0,-1)_\mu, \cr
                  (0,0,-2,0,0,1,\;\,1)_\nu \cr } \Bigg\rangle,    }
where $\lambda, \mu, \nu \not= 0$ and 
$(c_1,c_2,\cdots, c_7)_\lambda$ represents the $\IC^*$ action 
$x_i \mapsto \lambda^{c_i} x_i \; (\lambda\not=0)$. The quotient of 
$\IC^7$ by this group does not behave well, however if we remove some bad loci 
we can obtain a good manifold {\Cox}. In our case, the bad loci 
are determined by the toric data coming from the reflexive polytope and 
there are two different ways to make sense of the quotient by removing 
either the loci
\eqn\Zsub{
{\cal Z}=(x_1=x_2=x_3=0)\cup(x_4=x_5=0)\cup(x_6=x_7=0) \; \subset {\bf C}^7 
\;,}
or 
\eqn\ZsubFlop{
\tilde{\cal Z}=(x_1=x_2=x_4=x_5=0)\cup(x_3=x_7=0)\cup(x_4=x_5=x_6=0) 
\; \subset {\bf C}^7 
\;.}
Correspondingly we have toric varieties;
\eqn\Pabc{
{\bf P}_{\Sigma(a,b,c)} := \( {\bf C}^7 \setminus {\cal Z} \) / 
\( {\bf C}^* \)^3_{(a,b,c)} \;\;, \;\;
\tilde{\bf P}_{\Sigma(a,b,c)} := \( {\bf C}^7 \setminus \tilde{\cal Z} \) / 
\( {\bf C}^* \)^3_{(a,b,c)} \;\;. }
Looking into the detail of the relevant toric data, it is easy to see 
that these two toric varieties are related by the flop in two dimensions (, 
see e.g. {\MVI}{\KMV}). In the both cases the general section $s$ of the anti-
canonical bundle defines a Calabi-Yau hypersurface $X=s^{-1}(0)$. In what 
follows we will be mainly concerned with the hypersurface in 
$\tilde{\bf P}_{\Sigma(a,b,c)}$ because it contains a del Pezzo surface
as an toric divisor, which may be contracted. (The hypersurface in 
${\bf P}_{\Sigma(a,b,c)}$ contains a rational elliptic surface 
and K3 surface, and will be studied elsewhere. ) We denote the defining 
equation of the Calabi-Yau hypersurfaces by 
\eqn\Models{
\eqalign{
\tilde X_{E(8)}: & \;
x_1^2+x_2^3+x_3^6\{(x_4^6+x_5^6)x_6^{12}+(x_4^{18}+x_5^{18})x_7^{12} \}
=0 \; \subset \tilde \IP_{\S(3,2,1)}, \cr
\tilde X_{E(7)}: & \;
x_1^2+x_2^4+x_3^4\{(x_4^4+x_5^4)x_6^{8}+(x_4^{12}+x_5^{12})x_7^{8} \}
=0 \; \subset \tilde \IP_{\S(2,1,1)}, \cr
\tilde X_{E(6)}: & \;
x_1^3+x_2^3+x_3^3\{(x_4^3+x_5^3)x_6^{6}+(x_4^{9}+x_5^{9})x_7^{6} \}
=0 \; \subset \tilde \IP_{\S(1,1,1)}, \cr
}   }
where it is understood that all possible deformations, which 
are homogeneous polynomials with respect to the torus actions 
$\IC^*(a,b,c)$, are included in the defining equations. The 
homogeneous degree has been determined from the anti-canonical 
bundle (divisor) $-K_{{\bf P}_\S}$ of the ambient spaces. We see explicitly 
from the defining equation that the divisor 
$D_7=(x_7=0)$ restricted to $\tilde X_{E(k)}$ describes the 
del Pezzo surfaces $Bl_k$ for $k=6,7,8$.  Also the divisor $D_3=(x_3=0)$ 
restricted to $\tilde X_{E(k)}$ is $\IP^2$. (Hereafter we will abuse 
the notation $D_i$ to represent the corresponding divisor restricted 
to the hypersurface $\tilde X_{E(k)}$.) In our situation the del Pezzo 
surfaces appear  as the degree $d=a+b+c$ hypersurface in $\IP^3(a,b,c,1)$.

Our Calabi-Yau hypersurfaces are typical examples of those by Batyrev, 
and we may associate a pair of reflexive polytopes for mirror symmetry of 
these spaces. For example for the Calabi-Yau hypersurfaces 
$\tilde X_{E(k)} (k=6,7,8)$ we associate a pair of reflexive polytopes 
$(\D(a,b,c),\Ds(a,b,c))$ with vertices of $\Ds(a,b,c)$ given by 
$$
\eqalign{&
\ns{0}=(0,0,0,0) \;\,\; \ns{1}=(1,0,0,0) \;,\; \ns{2}=(0,1,0,0) \;,\; 
\ns{3}=(\rms a, \rms b,0,0) \;,\; 
\ns{4}=(0,0,0,1) \;,\; \cr
& \qquad
\ns{5}=(\rms 3 a, \rms 3 b, \rms 1, \rms 1) \;,\; 
\ns{6}=(0,0,1,0) \;,\; 
\ns{7}=(\rms 2a , \rms 2b, \rms 1,0) \;.\;  \cr}
$$
The polar duality of the polytope determines uniquely the polytope 
$\D(a,b,c)$.  Given the polytopes we may determine the Hodge numbers 
of the Calabi-Yau hypersurfaces to be $(h^{1,1}(X_\D),h^{2,1}(X_\D))=
(3,243), (4,148), (5,101)$ for $\tilde X_{E(8)}, \tilde X_{E(7)}$ and 
$\tilde X_{E(6)}$, respectively. 
Using the formula {\chow} for the Chow ring, we see that 
$H^{1,1}(X_\D,\IQ)$ has dimensions three for all $\tilde X_{E(k)} \; 
(k=6,7,8)$. The fact that we have less dimensions than those expected 
from $h^{1,1}(X_\D)$ for $\tilde X_{E(7)}$ and $\tilde X_{E(6)}$ 
means we have so-called twisted sectors for these models. There is a way 
to remedy this situation modifying the toric data slightly, however this 
is not important for our local mirror calculations. 

For the dual of the K\"ahler cone in $H^{1,1}(X_\D,\IQ)$, the Mori cone, 
we will find three generators (, see {\HLY} for detailed calculations,);
\eqn\basismori{
\eqalign{
&l^{(1)}=(\;\,0;  0, 0,\;\,1,   -1,   -1, \;0,\;\,1),  \cr
&l^{(2)}=(\;\,0;  0, 0  , -3,\;\,1,\;\,1, \;1,\;\,0),  \cr
&l^{(3)}=(   -d;  a, b,\;\,0,\;\,c,\;\,1, \;0,   -1). \quad  
\cr } }
These bases have their duals $J_1, J_2$ and $J_3$, which are related 
to the toric divisors $D_i$ in {\chow} by 
\eqn\JtoD{
D_7=J_1-J_3\;,\; D_3=J_1-3 J_2 \;,\; D_0 = -d J_3 \;\;, }
where $D_0=-(D_1+\cdots+D_7)$, the canonical class 
$K_{\tilde \IP_{\S(a,b,c)}}$.  The ring structure of 
$H^{even}_{toric}(X_\D,\IQ)$ may be recovered from the following data; 
\eqn\topData{
\eqalign{
& 
\tilde X_{E(6)} :  27 \; D_H^3 + 27\; D_3^3 + 3 \; D_7^3  \;\;, \;\; 
 (c_2, \vec D)=(90, -18, 6) \cr 
& 
\tilde X_{E(7)} :  18 \; D_H^3 + 18\; D_3^3 + 2 \; D_7^3  \;\;, \;\; 
 (c_2, \vec D)=(96, -12, 8) \cr 
& 
\tilde X_{E(8)} :  \;\, 9 \; D_H^3 \;+\,9\; D_3^3 \; + \,\; D_7^3  \;\;, \;\; 
 (c_2, \vec D)=(102, -6, 10) \cr 
} }
where we have used the convention that the coefficient of $D_i D_j D_k$ 
represents the corresponding cubic intersection number 
$\int_{X_\D} D_iD_jD_k$, and $(c_2,\vec D)=
( (c_2,D_H), (c_2,D_3), (c_2,D_7) )$ with $D_H:=J_1$.  

The above simple form of intersection numbers comes from 
the fact that the divisors $D_7$ and $D_3$ are isomorphic, respectively, 
to the del Pezzo surfaces $Bl_k \,(k=6,7,8)$ and $\IP^2$, 
which may be contracted to a point. In fact we 
see in {\basismori} that the rational curves $l^{(3)}$ and $l^{(2)}$ are  
contained, respectively, in the divisors $D_7$ and $D_3$. Thus 
$l^{(3)}$ and $l^{(2)}$ are the extremal rays for the respective contractions. 
In the next subsection we will study in detail the monodromy formula {\horja} 
for the contractions $l^{(3)}$ and $l^{(2)}$.

\subsec{ Local limit of the monodromy --- 
del Pezzo surfaces $ \IP^2, Bl_6, Bl_7, Bl_8$ }

Now we are ready to consider  the monodromy problem under 
the local mirror symmetry limit to the del Pezzo surfaces. 
We will focus on the primitive contractions corresponding the extremal 
rays $l^{(2)}$ and $l^{(3)}$, explicit in our models $\tilde X_{E(k)}$. 

Specifying a contraction, say $l^{(3)}$, we have a subvariety 
which is isomorphic to $\IP^1$ in the compactification of the moduli 
space of the complex structures of $\tilde X_{E(k)}^\vee$. Namely the 
local coordinate of the $\IP^1$ about the LCSL is given by 
$z=(-1)^{l_0^{(3)}}a^{l^{(3)}}$ and the $\IP^1$ is located at $x=y=0 
\;( x=(-1)^{l_0^{(1)}}a^{l^{(1)}}, y=(-1)^{l_0^{(2)}}a^{l^{(2)}})$. This 
is exactly the situation in which we take the local mirror symmetry 
limit to del Pezzo surfaces{\CKYZ}. If we take $l^{(2)}$, then we will 
come to the local mirror symmetry limit to $\IP^2$. In both cases 
the general formula {\horja} due to Horja applies. The 
Picard-Fuchs equations (${\cal L}w=0$) which describe our local mirror 
symmetry limits also follow from the data of the extremal ray $l$, 
and they are given by,
\eqn\PFdelP{
\eqalign{
& 
\IP^2: {\cal L}=\{ \theta_y^2+3y(3\theta_y+2)(3\theta_y+1) \}\theta_y 
\;\;,\;\;
Bl_6: {\cal L}=\{ \theta_z^2+3z(3\theta_z+2)(3\theta_z+1) \}\theta_z  \cr
&
Bl_7: {\cal L}=\{ \theta_z^2+4z(4\theta_z+3)(4\theta_z+1) \}\theta_z 
\;\;,\;\;
Bl_8: {\cal L}=\{ \theta_z^2+12z(6\theta_z+5)(6\theta_z+1) \}\theta_z \;.
}}
We see from the form of the differential operators, the monodromy of the 
solutions is reducible for all cases, which will be interpreted later.

Now let us look closely our hypergeometric series {\IIaIIb} under the local 
mirror symmetry limit to the del Pezzo surfaces $Bl_6, Bl_7, Bl_8$.  
In the IIB side the 
local mirror symmetry limit means the limit  $x,y\rightarrow 0$ in the 
hypergeometric series. From our definition of the hypergeometric series 
combined with the ring structure of $H^{even}_{toric}(X_\D,\IQ)$, it is 
straightforward to obtain the following form of the limit;
\eqn\limwoDi{
\eqalign{
&w_0\left( \vec x, {\vec J \over 2\pi i} \right) {\rm mod} \; Ann \, D_7
 \Big|_{x,y \rightarrow 0}  = \cr
&{\bf 1} + D_7 \(-{1\over 2\pi i} \pd_{\rho_3} w_0(z) \) 
+ D_7^2 \({1\over2}{1\over (2\pi i)^2} \pd_{\rho_3}^2 w_0(z) \) 
+ D_7^3 \(-{1\over6}{1\over (2\pi i)^3} \pd_{\rho_3}^3 w_0(z) \) \;\;, \cr
} }
where $Ann \, D_7 = \{ v \in H^{even}_{toric}(X_{\Delta},{\bf Q}) \;|\; 
D_7 \cdot v =0 \;\}\otimes {\bf C}[{\rm log} x]\{x\}$. 
Since the first three terms should give the local solutions of the 
3rd order differential equations {\PFdelP}, we arrange the above 
formal expansion to $w_0(z,{J\over 2\pi i})=
({\bf 1}, D_7, D_7^2).\Pi_{B, local}(x)$ neglecting the $D_7^3$ term.  
In case of $\IP^2$ we expand the series {\IIaIIb} via 
the basis $\Pi^{A,local}=({\bf 1}, D_3, D_3^2)$, and correspondingly 
we have $\Pi_{B, local}(y)=\;^t(1,-{1\over3}\pd_{\rho_2}w_0, 
{1\over18}\pd^2_{\rho_2} w_0 )$.

In Table 3, we have listed the monodromy matrices for the hypergeometric 
series $\Pi_{B,local}$. Though the the evaluations of the monodromy is 
straightforward, we present, for reader's convenience, our definitions  
of the hypergeometric series about $z=\infty$ and its connection matrix 
$N$ to the series about $z=0$. We should note that the monodromy about 
$z=0,\infty$ is easy to be determined.  For $Bl_k$  
we have defined the hypergeometric series about $z=\infty$  
for $j=0,1$, in addition to the obvious one $1$, by 
$$
\eqalign{
w^\infty_j(z)&=\sum_{k=1}^{d-1}
        (1-\alpha^{ak})(1-\alpha^{bk})(1-\alpha^{ck})\alpha^{kj} 
        \tilde w_k(z) \;\;, \cr
\tilde w_k(z) & =
{1\over d}{1\over (2\pi i)^3} \sum_{N=0}^\infty 
{ \Gamma(a N+{a k\over d})\Gamma(b N+{b k \over d}) \Gamma(N+{k \over d})^2 
   \over 
  \Gamma(d N+k)\Gamma(1+N+{k\over d}) } (-z)^{-N-{k \over d}}  \;\;,  \cr}
$$
with $\alpha$ being the primitive root of $\alpha^d=1$ $(d=a+b+c)$. 
The connection matrix has determined by analytic continuation of the 
above series back to the region $z=0$. ($\IP^2$ has the same hypergeometric 
series about $z=\infty$ as $Bl_6$, but differs about $z=0$ because the 
$l$ vector is different. This simply results in the difference in the 
connection matrix and also the monodromy matrices.)

$$
\def\vspace#1{ \omit &height #1 &  \omit&&  \omit && \omit && \omit &\cr }
\vbox{\offinterlineskip 
\halign{
\strut#&\vrule#&  $\;$ \hfil # \hfil  
&\vrule#&  \hfil # \hfil  
&\vrule#&  \hfil # \hfil  
&\vrule#&  \hfil # \hfil  
&\vrule#&  \hfil # \hfil  
&\vrule#   \cr  
\noalign{\hrule}
&&
&& $\IP^2$ 
&& $Bl_6$ 
&& $Bl_7$ 
&& $Bl_8$ 
& \cr
\noalign{\hrule}
\vspace{1pt}
\noalign{\hrule}
&& $N$ 
&& $\(\matrix{ 1 & 0 & 0 \cr {1\over3} & -{2\over9} & -{1\over9} \cr 
              {5\over54} & -{2\over27} & -{1\over54} \cr}\)$
&& $\(\matrix{ 1 & 0 & 0 \cr 1 & -{2\over3} & -{1\over3} \cr 
              {1\over2} & -{2\over3} & -{1\over6} \cr}\)$
&&  $\(\matrix{ 1 & 0 & 0 \cr 1 & -{1\over2} & -{1\over2} \cr 
              {1\over2} & -{3\over4} & -{1\over4} \cr}\)$
&& $\(\matrix{ 1 & 0 & 0 \cr 1 &  0  & -1 \cr 
              {1\over2} & -1 & -{1\over2} \cr}\)$
&\cr
\noalign{\hrule}
&& $M_0$ 
&& $\(\matrix{ 1 & 0 & 0 \cr -{1 \over 3} & 1 &  0 \cr 
              {1\over18} & -{1\over3} &  {1} \cr}\)$
&& $\(\matrix{ 1 & 0 & 0 \cr -{1} & 1 &  0 \cr 
              {1\over2} & -{1} &  {1} \cr}\)$
&& $\(\matrix{ 1 & 0 & 0 \cr -{1} & 1 &  0 \cr 
              {1\over2} & -{1} &  {1} \cr}\)$
&&  $\(\matrix{ 1 & 0 & 0 \cr -{1} & 1 &  0 \cr 
              {1\over2} & -{1} &  {1} \cr}\)$
& \cr
&& $M_1$
&& $\(\matrix{ 1 & 0 & 0 \cr 0 & -{1\over2} &  9 \cr 
               0 & -{1\over4} &  {5\over2} \cr}\)$
&& $\(\matrix{ 1 & 0 & 0 \cr 1 & -{1\over2} &  3 \cr 
               {1\over2} & -{3\over4} &  {5\over2} \cr}\)$
&& $\(\matrix{ 1 & 0 & 0 \cr 1 & 0 &  2 \cr 
               {1\over2} & -{1\over2} &  {2} \cr}\)$
&&  $\(\matrix{ 1 & 0 & 0 \cr 1 & {1\over2} &  1 \cr 
               {1\over2} & -{1\over4} &  {3\over2} \cr}\)$
& \cr
\noalign{\hrule}
}}
$$
{\leftskip1cm \rightskip1cm \noindent
{\bf Table 3}.  The connection matrix and the monodromy matrices for del 
Pezzo surfaces. The connection matrix relates the hypergeometric series 
by $\Pi_{B,local}=N. \,^t(1,w^\infty_0,w^\infty_1)$.  \par}

\vskip0.6cm

Corresponding to this hypergeometric series we consider the type IIA 
monodromy acting on the row vector $\Pi^{A, local}=
({\bf 1}, D_7, D_7^2)$. Up to conjugation, the formula {\horja} for the 
type IIA monodromy may be identified with the monodromy about the discriminant 
$dis=1+27 z, 1+64 z, 1+432 z$, for $Bl_6, Bl_7, Bl_8$, respectively. 
In these cases the general formula {\horja} may be written 
\eqn\Tiii{
{\cal T}_3: \gamma \mapsto \gamma-(1-{\rm e}^{D_7}) 
\int_{C_\xi}{(1-{\rm e}^{-d \xi}) \over 
             (1-{\rm e}^{-a \xi})(1-{\rm e}^{-b \xi}) 
             (1-{\rm e}^{{2 D_H +D_6 \over 3}-\xi})^2} \gamma(\xi) 
       {d\xi \over 2\pi i} \;\;,
}
where the residue is taken about $\xi=0$ and $\xi={2 D_H +D_6 \over 3}$. 
When writing this formula we use the rational equivalences, 
$D_3=a J_3, D_2=b J_3, D_4=D_5=-J_1+J_2+J_3=-{2 D_H +D_6 \over 3}+J_3$ 
and $D_0=-d J_3$. Also we use $D_7=D_H-J_3$ and the intersection 
numbers {\topData} when evaluating $\gamma(\xi)$ for 
$\gamma={\bf 1}, D_7, D_7^2$.  For the $\IP^2$ contraction 
the corresponding formula is more simple and has the following form,
\eqn\Tii{
{\cal T}_2: \gamma \mapsto \gamma-(1-{\rm e}^{D_3}) 
\int_{C_\xi} { 1 \over 
             (1-{\rm e}^{-\xi})^3 } \gamma(\xi) 
       {d\xi \over 2\pi i} \;\;. 
}
Here we use $D_3=D_H-3 J_2$ for the determination $\gamma(\xi)$ for $\gamma=
{\bf 1}, D_3, D_3^2$.

\vskip0.6cm

In both cases it is easy to verify that the type IIA monodromy 
reproduces, up to conjugations, the monodromy $M_1$ 
about the discriminant for the hypergeometric series $\Pi_{B, local}$. 
Precisely for $Bl_k\; (k=6,7,8)$ we verify that the monodromy operation 
${\cal T}_2$ on the basis $\Pi^{A,local}$ exactly reproduces the matrix 
$M_1$ in Table 3, and for $\IP^2$ it coincides with $M_0^{-1}M_1 M_0$ in 
Table 3.  This is not merely a verification of the general 
results obtained in {\Horja}, {\it but} our point here is that we 
may arrange the expansion {\limwoDi} using the pairing in ``Theorem'' 2. 
Namely we may arrange the hypergeometric series, for example, 
into the following,
\eqn\basisInt{
\eqalign{
&w_0\left(\vec x,{\vec J \over 2\pi i} \right) \; {\rm mod} \; Ann \, D_7 
\Big|_{x,y\rightarrow 0}  = \cr
& w_0(z){\bf 1}+
w^{(1)}(z)\left( D_7 -{1\over12}c_2(X_{\Delta})D_7-{1\over2}D_7^2 \right)
+ w^{(2)}(z){1\over h}D_7^2 + w^{(3)}(z) \left( -{1\over h} D_7^3 \right)
\quad,
} }
where we set $h=1,2,3$, respectively, for $Bl_8, Bl_7, Bl_6$, and also   
note that $w_0(z), w_1(z), w_2(z)$ are linear combinations of 
$1, \pd_{\rho_3} w_0(z), \pd^2_{\rho_3} w_0(z)$. 
In the arrangement above {\basisInt}, we have evaluated the Chern character 
of the ideal sheaf ${\cal I}_{p}$ on the del Pezzo surface 
$i: Bl_k \hookrightarrow X_\D$;   
\eqn\idelaBl{
\ch(i_* {\cal I}_{p})= D_7 - {1\over 2} D_7^2 -{1\over12}c_2(X) D_7 \;\;, 
}
using $c_1({\cal I}_{p})=0, ch_2({\cal I}_p)=- \Vol_S$ 
for the ideal sheaf and $c_1(D_7)=-D_7$ which follows from the adjunction 
formula for $0 \rightarrow T_S \rightarrow T_X \rightarrow N \rightarrow 0$ 
with $c_1(N)|_S=D_7$ and $c_1(X)=0$. We also use 
${1\over12}(c_1(S)^2+c_2(S))= \Vol_S$ for rational surfaces $S$,  
($\chi(S,{\cal O}_S)=1$). 
As for the 3-rd term of {\basisInt}, $w^{(2)}\times {1\over h}D_7^2$, 
we identify this with some sheaf ${\cal E}$, tensored with a suitable 
line bundle on it, supported on the 
canonical divisor $-c_1(D_7)=D_7$ of the surface (see {\identSheaves}). 
These two sheaves should correspond to some 
3-cycles in the mirror $X^\vee$, which are not explicit in our argument 
(, see the discussions in the next section for recent works on this). 
We remark that the reducibility of the monodromy is evident 
in our interpretation because the D-brane charge ${\bf 1}$ is not local but 
${\bf 1}=\ch({\cal O}_X)$.

As we see in Table 4, after arranging the expansion of the hypergeometric 
series to {\basisInt}, the monodromy matrices becomes integral. This 
means that the sheaves we looked above may be part of the integral 
generators of $K_{hol}(X)$. In  case of $\IP^2$, however, the monodromy 
matrix contains one rational number ${1\over 3}$. This should be understood  
by the fact that the anti-canonical class $c_1(D_3)$ of $\IP^2$ is not 
a primitive class but three times of the line on it.

$$
\def\vspace#1{ \omit &height #1 &  \omit&&  \omit && \omit && \omit &\cr }
\vbox{\offinterlineskip 
\halign{
\strut#&\vrule#&  $\;$ \hfil # \hfil  
&\vrule#&  \hfil # \hfil  
&\vrule#&  \hfil # \hfil  
&\vrule#&  \hfil # \hfil  
&\vrule#&  \hfil # \hfil  
&\vrule#   \cr  
\noalign{\hrule}
&&
&& $\IP^2$ 
&& $Bl_6$ 
&& $Bl_7$ 
&& $Bl_8$ 
& \cr
\noalign{\hrule}
\vspace{1pt}
\noalign{\hrule}
&& $M_0$ 
&& $\(\matrix{  1 &  0 & 0 \cr   
       -{1\over3} &  1 & 0  \cr 
               -1 & -3 &  1\cr}\)$
&&  $\(\matrix{ 1 & 0 & 0  \cr 
               -1 & 1 & 0  \cr 
                0 &-3 & 1  \cr}\)$
&&  $\(\matrix{ 1 & 0 & 0  \cr 
               -1 & 1 & 0  \cr 
                0 &-2 & 1  \cr}\)$
&&  $\(\matrix{ 1 & 0 & 0  \cr 
               -1 & 1 & 0  \cr 
                0 &-1 & 1  \cr}\)$
& \cr
&& $M_1$
&& $\(\matrix{  1 & 0 & 0  \cr   
                0 &-5 & 1  \cr 
                0 &-36& 7  \cr}\)$
&&  $\(\matrix{ 1 & 0 & 0 \cr 
                1 &-2 & 1  \cr 
                3 &-9 & 4  \cr}\)$
&&  $\(\matrix{ 1 & 0 & 0 \cr 
                1 &-1 & 1  \cr 
                2 &-4 & 3  \cr}\)$
&&  $\(\matrix{ 1 & 0 & 0 \cr 
                1 & 0 & 1  \cr 
                1 &-1 & 2  \cr}\)$
& \cr
\noalign{\hrule}
}}
$$
{\leftskip1cm \rightskip1cm \noindent
{\bf Table 4}. Integral monodromy matrices for $S=Bl_k \; (k=6,7,8)$. 
We identify the corresponding sheaves with ${\cal O}_X, i_*{\cal I}_p, 
i_*{\cal E}$ in $D(X)$.  Monodromy matrices of $\IP^2$ and 
$Bl_k$ have also been calculated, respectively, in {\DFR} and {\LMW}. 
 \par}

\newsec{ Discussions }

We have proposed the monodromy invariant pairing {\IinvP} between the D-brane 
charges. As we have argued, this pairing has its ground on the homological 
mirror symmetry due to Kontsevich. It is interesting to note that the pairing 
was essentially utilized when evaluating the prepotential in the very 
beginning of the mirror symmetry and its application to Gromov-Witten 
invariants.

In the following we summarize related subjects which we haven't addressed in 
this article.

In this paper we have restricted our attentions to the primitive 
contractions, where the extremal rays or equivalently some edges in the 
secondary polytope play their roles to specify the torus invariant 
orbit $\IP^1$ in the complex structure moduli space. 
Zariski theorem of Lefschetz type for hypersurfaces says that 
for a generic line $\IP^1$, generic to the discriminant variety, 
there is a surjective homomorphism 
$\pi_1(\IP^1 \setminus (Dis \cap \IP^1)) 
\rightarrow \pi_1(\bar {\cal M} \setminus Dis)$ for the deformation space 
$\bar {\cal M}$.  It is not clear the analysis restricting to the torus 
invariant $\IP^1$ suffices to find all necessary generators for the monodromy 
group, although for the one, two parameter examples we looked in this paper 
we verify that the monodromy group is generated by them. 

The Calabi-Yau hypersurfaces we have looked in the last section admit 
topology changes due to the flop operations. According to {\AGM} the 
(complexified) K\"ahler moduli spaces of topologically different Calabi-Yau 
manifolds $X$ and $X'$ are unified in the complex structure moduli space 
of their mirror $X^\vee$ by analytic continuation. Combined with the 
homological mirror symmetry, we should have the categorical equivalence 
$D(X)\cong D(X')$ for their derived categories. The full picture of the 
moduli spaces of our models in this respect will be reported elsewhere.

Finally we address to recent progresses in physics. The homological 
mirror symmetry due to Kontsevich may be interpreted 
as the mirror symmetry of D-branes in string theory. 
In ref.{\OOY} the conformal field theory analogue (boundary 
states) of the D-branes was first formulated, and more recently in the 
work {\BDoug} the relations between the geometry of the cycles and 
the boundary states  of conformal filed theory (at Geppnar point) have 
been pursued. In this respect the two parameter models 
we looked in the text have also been  analyzed in refs.{\DR}{\KLLW}{\Sch}. 
The local mirror symmetry limit of these D-brane analysis considered 
in {\DFR}. Especially the local mirror symmetry limit and also 
the monodromy problem have been studied in general using Landau-Ginzburg 
theory in a recent paper{\HIV}, which 
appeared when this article was being completed. The paper {\HIV} proposes 
a way to construct 3-cycles which corresponds to sheaves on (toric) del Pezzo 
surfaces, which we haven't looked in this paper. Our pairing contained 
in the formal expansion of the hypergeometric series{\IIaIIb}{\homCD} 
provides us the corresponding hypergeometric series 
to a sheaf without specifying the 3-cycle explicitly, and seems to provide 
a complementary approach to {\HIV}.  
We will come to this problem in future investigations.


\vskip1cm

\appendix{A}{Brief summary of the monodromy calculations}

In this appendix, following {\CdGP}, we briefly summarize the 
monodromy calculations done in the text. Here we write the case of 
the quartic hypersurface in $\IP^3$, the extension to other cases 
are straightforward. As described in section 2, the relevant hypergeometric 
series about a LCSL are determined by the toric data $l=(-4;1,1,1,1)$, 
from which we have $w_0(x)=\sum_{n=0}^{\infty} {(4n)! \over (n!)^4} x^n$. 
The analytic continuation to $x=\infty$ may be done by the Barnes integral 
representation 
\eqn\appOne{
\eqalign{
w_0(x)&=\sum_{n=0}^{\infty} {(4n)! \over (n!)^4} x^n \cr 
& = {1\over 2\pi i} \int_C ds \Gamma(-s) 
  { \Gamma(4s+1) (-1)^s \over \Gamma(s+1)^3} x^s \cr
& = {1\over 2\pi i} \int_C ds \Gamma(4s+1) 
  { \Gamma(-s) (-1)^s \over \Gamma(s+1)^3} x^s  \cr
& = -{1\over 4} \sum_{m=1}^\infty { \Gamma({m\over4}) \over 
  \Gamma(m) \Gamma(1-{m\over4})^3 } (-1)^{3 m\over 4} x^{-{m\over4}} \;\;,
\cr} }
where in the second line the contour $C$ encircle the poles at 
$s=0,1,2,\cdots$ counterclockwise, while in the third line it is 
deformed to encircle the poles at $s=-{m\over 4} (m=1,2,\cdots)$ clockwise. 
This completes the analytic continuation 
to $x=\infty$. Other hypergeometric series about $x=\infty$ are defined 
by
\eqn\appTwo{
w_j^\infty(x)=-{1\over4}\sum_{m=1}^\infty 
{ \Gamma({m\over4}) \over 
  \Gamma(m) \Gamma(1-{m\over4})^3 } \alpha^{3 m\over 2} \alpha^{m j}
x^{-{m\over4}}  \;\; (j=0,1,2,3) \;\;, }
where $\alpha$ is the primitive root of $\alpha^4=1$. Setting $m=4 N + k$ and 
using the relation $\Gamma(z)\Gamma(1-z)={\pi \over {\rm sin} \pi z }$, we may 
rewrite {\appTwo} as
\eqn\appThree{
w_j^\infty(x)=\sum_{k=1}^4 (1-\alpha^k)\alpha^{kj} \tilde w_k(x) \;\;,\;\;
\tilde w_k(x)={1\over 4}{1\over (2\pi i)^3} \sum_{N=0}^\infty 
  {\Gamma(N+{k\over4})^4 \over \Gamma(4N+k) } x^{-N-{k\over4}} \;\;. }
These four series are not independent but have one relation 
$w_0^\infty + w_1^\infty + w_2^\infty + w_3^\infty=0$. 
For the analytic continuation back to $x=0$ we utilize again the following 
integral representation for $\tilde w_k(x)$;
\eqn\appFour{
\eqalign{
\tilde w_k(x)
&=\int_{C_1} {ds \over {\rm e}^{2\pi i}-1} {1\over 4} 
{1\over (2\pi i)^3} {\Gamma(s+{k\over4})^4 \over \Gamma(4s+k) } 
x^{-s-{k\over4}} \cr
&=-\int_{C_1'} 
{ds \over {\rm e}^{2\pi i}-1} {1\over 4} 
{1\over (2\pi i)^3} {\Gamma(s+{k\over4})^4 \over \Gamma(4s+k) } 
x^{-s-{k\over4}} \;\;,\cr }}
where in the first line $C_1$ encircles the poles at $s=0,1,2,\cdots$ 
counterclockwise, while in the second line it is deformed to $C_1'$ which 
encircles the higher 3rd order poles at $s=-m-{k\over4} \; (m=0,1,2,\cdots)$ 
counterclockwise. The higher order poles above introduce the logarithms 
about $x=0$, up to $(\log x)^2$, which should be connected to the logarithmic 
singularity in $w_0, \pd_\rho w_0, \pd_\rho^2 w_0$ at our LCSL. The connection 
matrix may be determined simply by comparing our solutions at LCSL and 
the analytic continuation of $w_j^\infty(x)$. The monodromy at LCSL simply 
comes from the logarithms in the hypergeometric series, and easily determined. 
Also the monodromy about $x=\infty$ is easily determined. This is because 
the monodromy is represented in the basis $w_j^\infty(x)$ by 
$w_0^\infty \rightarrow w_1^\infty, 
w_1^\infty \rightarrow w_2^\infty, 
w_2^\infty \rightarrow w_3^\infty=-w_0^\infty-w_1^\infty-w_2^\infty$ using 
the relation noted above. 

Monodromy matrices and the connection matrix appeared in the text are 
determined in this way.

\vskip0.5cm

{\bf References }

\item{\AGM} \refAGM
\item{\Bat} \refBat
\item{\BO} \refBO
\item{\Bri} \refBri
\item{\BDoug} \refBDoug
\item{\CdGP} \refCdGP
\item{\CdTwoI} \refCdTwoI
\item{\CdTwoII} \refCdTwoII
\item{\CKYZ} \refCKYZ
\item{\Cox} \refCox
\item{\DR} \refDR
\item{\Dol} \refDol
\item{\DFR} \refDFR
\item{\Ful} \refFul
\item{\Gi}  \refGi
\item{\HIV} \refHIV
\item{\Horja} \refHorja
\item{\Hos} \refHos
\item{\HKTYI} \refHKTYI
\item{\HKTYII} \refHKTYII
\item{\HLY} \refHLY
\item{\HLYII} \refHLYII
\item{\HLYm} \refHLYm
\item{\KLLW} \refKLLW
\item{\KLM} \refKLM
\item{\KMV} \refKMV
\item{\KT} \refKT
\item{\Kob} \refKob
\item{\KoI} \refKoI
\item{\KoII} \refKoII
\item{\LMW} \refLMW
\item{\LLY} \refLLY
\item{\LY} \refLY
\item{\Mor} \refMor
\item{\MVI} \refMVI
\item{\MVII} \refMVII
\item{\MukaiI} \refMukaiI
\item{\MukaiII} \refMukaiII
\item{\Oda} \refOda
\item{\OOY} \refOOY
\item{\Sch} \refSch
\item{\Sti} \refSti
\item{\Str} \refStr
\item{\SYZ} \refSYZ
\item{\Vafa} \refVafa

\bye